\input harvmac.tex
\newdimen\tableauside\tableauside=1.0ex
\newdimen\tableaurule\tableaurule=0.4pt
\newdimen\tableaustep
\def\phantomhrule#1{\hbox{\vbox to0pt{\hrule height\tableaurule width#1\vss}}}
\def\phantomvrule#1{\vbox{\hbox to0pt{\vrule width\tableaurule height#1\hss}}}
\def\sqr{\vbox{%
  \phantomhrule\tableaustep
  \hbox{\phantomvrule\tableaustep\kern\tableaustep\phantomvrule\tableaustep}%
  \hbox{\vbox{\phantomhrule\tableauside}\kern-\tableaurule}}}
\def\squares#1{\hbox{\count0=#1\noindent\loop\sqr
  \advance\count0 by-1 \ifnum\count0>0\repeat}}
\def\tableau#1{\vcenter{\offinterlineskip
  \tableaustep=\tableauside\advance\tableaustep by-\tableaurule
  \kern\normallineskip\hbox
    {\kern\normallineskip\vbox
      {\gettableau#1 0 }%
     \kern\normallineskip\kern\tableaurule}%
  \kern\normallineskip\kern\tableaurule}}
\def\gettableau#1 {\ifnum#1=0\let\next=\null\else
  \squares{#1}\let\next=\gettableau\fi\next}

\tableauside=1.0ex
\tableaurule=0.4pt
\input epsf
\noblackbox

\def\tpsi{{\widetilde \psi}}

\def\tb{{\bar r}}
\def\ra{{\rightarrow}}
\def\og{{\overline G}}



\def\unlockat{\catcode`\@=11}
\def\lockat{\catcode`\@=12}

\unlockat

\def\newsec#1{\global\advance\secno by1\message{(\the\secno. #1)}
\global\subsecno=0\global\subsubsecno=0\eqnres@t\noindent
{\bf\the\secno. #1}
\writetoca{{\secsym} {#1}}\par\nobreak\medskip\nobreak}
\global\newcount\subsecno \global\subsecno=0
\def\subsec#1{\global\advance\subsecno
by1\message{(\secsym\the\subsecno. #1)}
\ifnum\lastpenalty>9000\else\bigbreak\fi\global\subsubsecno=0
\noindent{\it\secsym\the\subsecno. #1}
\writetoca{\string\quad {\secsym\the\subsecno.} {#1}}
\par\nobreak\medskip\nobreak}
\global\newcount\subsubsecno \global\subsubsecno=0
\def\subsubsec#1{\global\advance\subsubsecno
\message{(\secsym\the\subsecno.\the\subsubsecno. #1)}
\ifnum\lastpenalty>9000\else\bigbreak\fi
\noindent\quad{\secsym\the\subsecno.\the\subsubsecno.}{#1}
\writetoca{\string\qquad{\secsym\the\subsecno.\the\subsubsecno.}{#1}}
\par\nobreak\medskip\nobreak}

\def\subsubseclab#1{\DefWarn#1\xdef
#1{\noexpand\hyperref{}{subsubsection}%
{\secsym\the\subsecno.\the\subsubsecno}%
{\secsym\the\subsecno.\the\subsubsecno}}%
\writedef{#1\leftbracket#1}\wrlabeL{#1=#1}}
\lockat

\def\IL{{\relax{\rm I\kern-.18em L}}}
\def\IH{{\relax{\rm I\kern-.18em H}}}
\def\IR{{\relax{\rm I\kern-.18em R}}}
\def\IE{{\relax{\rm I\kern-.18em E}}}
\def\IC{{\relax\hbox{$\inbar\kern-.3em{\rm C}$}}}
\def\IZ{{\relax\ifmmode\mathchoice
{\hbox{\cmss Z\kern-.4em Z}}{\hbox{\cmss Z\kern-.4em Z}}
{\lower.9pt\hbox{\cmsss Z\kern-.4em Z}}
{\lower1.2pt\hbox{\cmsss Z\kern-.4em Z}}\else{\cmss Z\kern-.4em
Z}\fi}}
\def\CM {{\cal M}}

\def\CD {{\cal D}}
\def\CF {{\cal F}}

\def\CP {{\cal P }}

\def\CV {{\cal V}}
\def\CO {{\cal O}}
\def\CZ {{\cal Z}}

\def\CH {{\cal H}}
\def\CC {{\cal C}}
\def\CB {{\cal B}}

\def\CA{{\cal A}}

\def\CM {{\cal M}}

\def\CO {{\cal O}}

\def\CP {{\cal P }}
\def\CQ {{\cal Q }}

\def\CV{{\cal V }}
\def\CZ {{\cal Z }}

\font\manual=manfnt \def\dbend{\lower3.5pt\hbox{\manual\char127}}

\def\IZ{{\relax\ifmmode\mathchoice
{\hbox{\cmss Z\kern-.4em Z}}{\hbox{\cmss Z\kern-.4em Z}}
{\lower.9pt\hbox{\cmsss Z\kern-.4em Z}}
{\lower1.2pt\hbox{\cmsss Z\kern-.4em Z}}\else{\cmss Z\kern-.4em
Z}\fi}}

\def\CM {{\cal M}}

\def\CO {{\cal O}}

\def\CP {{\cal P }}
\def\CQ {{\cal Q }}

\def\CV{{\cal V }}
\def\CZ {{\cal Z }}

\def\CX{{\cal X}}

\def\om{{\overline M}}

\def\of{{\overline F}}
\def\delbar{{\overline \partial}}


\def\IZ{{\relax\ifmmode\mathchoice
{\hbox{\cmss Z\kern-.4em Z}}{\hbox{\cmss Z\kern-.4em Z}}
{\lower.9pt\hbox{\cmsss Z\kern-.4em Z}}
{\lower1.2pt\hbox{\cmsss Z\kern-.4em Z}}\else{\cmss Z\kern-.4em
Z}\fi}}
\def\IB{{\relax{\rm I\kern-.18em B}}}
\def\IC{{\relax\hbox{$\inbar\kern-.3em{\rm C}$}}}
\def\ID{{\relax{\rm I\kern-.18em D}}}
\def\IE{{\relax{\rm I\kern-.18em E}}}
\def\IF{{\relax{\rm I\kern-.18em F}}}
\def\IG{{\relax\hbox{$\inbar\kern-.3em{\rm G}$}}}
\def\IGa{{\relax\hbox{${\rm I}\kern-.18em\Gamma$}}}
\def\IH{{\relax{\rm I\kern-.18em H}}}
\def\II{{\mathchoice {\rm 1\mskip-4mu l} {\rm 1\mskip-4mu l}
          {\rm 1\mskip-4.5mu l} {\rm 1\mskip-5mu l}}}

\def\IK{{\relax{\rm I\kern-.18em K}}}
\def\IP{{\relax{\rm I\kern-.18em P}}}

\def\inbar{\,\vrule height1.5ex width.4pt depth0pt}

\font\cmss=cmss10 \font\cmsss=cmss10 at 7pt
\def\IR{\relax{\rm I\kern-.18em R}}
\def\IT{\relax{\rm I\kern-.45em T}}


\def\boxit#1{\vbox{\hrule\hbox{\vrule\kern8pt
\vbox{\hbox{\kern8pt}\hbox{\vbox{#1}}\hbox{\kern8pt}}
\kern8pt\vrule}\hrule}}
\def\mathboxit#1{\vbox{\hrule\hbox{\vrule\kern8pt\vbox{\kern8pt
\hbox{$\displaystyle #1$}\kern8pt}\kern8pt\vrule}\hrule}}


\def\inbar{\,\vrule height1.5ex width.4pt depth0pt}

\font\cmss=cmss10 \font\cmsss=cmss10 at 7pt
\def\IR{\relax{\rm I\kern-.18em R}}

\def\tb{{\bar t}}


\def\ra{{\longrightarrow}}

\let\includefigures=\iftrue
\newfam\black
\includefigures

\input epsf
\def\plb#1 #2 {Phys. Lett. {\bf B#1} #2 }
\long\def\del#1\enddel{}
\long\def\new#1\endnew{{\bf #1}}
\let\<\langle \let\>\rangle

\def\figin{\epsfcheck\figin}\def\figins{\epsfcheck\figins}
\def\epsfcheck{\ifx\epsfbox\UnDeFiNeD
\message{(NO epsf.tex, FIGURES WILL BE IGNORED)}
\gdef\figin##1{\vskip2in}\gdef\figins##1{\hskip.5in} blank space instead
\else\message{(FIGURES WILL BE INCLUDED)}
\gdef\figin##1{##1}\gdef\figins##1{##1}\fi}
\def\DefWarn#1{}
\def\figinsert{\goodbreak\midinsert}
\def\ifig#1#2#3{\DefWarn#1\xdef#1{fig.~\the\figno}
\writedef{#1\leftbracket fig.\noexpand~\the\figno}
\figinsert\figin{\centerline{#3}}\medskip
\centerline{\vbox{\baselineskip12pt
\advance\hsize by -1truein\noindent
\footnotefont{\bf Fig.~\the\figno:} #2}}
\bigskip\endinsert\global\advance\figno by1}
\else
\def\ifig#1#2#3{\xdef#1{fig.~\the\figno}
\writedef{#1\leftbracket fig.\noexpand~\the\figno}
\figinsert\figin{\centerline{#3}}\medskip
\centerline{\vbox{\baselineskip12pt
\advance\hsize by -1truein\noindent
\footnotefont{\bf Fig.~\the\figno:} #2}}
\bigskip\endinsert
\global\advance\figno by1}
\fi

\input xy
\xyoption{all}
\font\cmss=cmss10 \font\cmsss=cmss10 at 7pt
\def\inbar{\,\vrule height1.5ex width.4pt depth0pt}
\def\IC{{\relax\hbox{$\inbar\kern-.3em{\rm C}$}}}
\def\IP{{\relax{\rm I\kern-.18em P}}}
\def\IF{{\relax{\rm I\kern-.18em F}}}
\def\IZ{\relax\ifmmode\mathchoice
{\hbox{\cmss Z\kern-.4em Z}}{\hbox{\cmss Z\kern-.4em Z}}
{\lower.9pt\hbox{\cmsss Z\kern-.4em Z}}
{\lower1.2pt\hbox{\cmsss Z\kern-.4em Z}}\else{\cmss Z\kern-.4em
Z}\fi}
\def\IR{{\relax{\rm I\kern-.18em R}}}
\def\IQ{\relax\hbox{\kern.25em$\inbar\kern-.3em{\rm Q}$}}

\def\pmb#1{\setbox0=\hbox{#1}%
 \kern-.025em\copy0\kern-\wd0
 \kern.05em\copy0\kern-\wd0
 \kern-.025em\raise.0433em\box0 }
\font\cmss=cmss10
\font\cmsss=cmss10 at 7pt
\def\rlx{\relax\leavevmode}
\def\Cop{\relax\,\hbox{$\inbar\kern-.3em{\rm C}$}}
\def\Rop{\relax{\rm I\kern-.18em R}}
\def\Nop{\relax{\rm I\kern-.18em N}}
\def\Pop{\relax{\rm I\kern-.18em P}}
\def\Zop{\rlx\leavevmode\ifmmode\mathchoice{\hbox{\cmss Z\kern-.4em Z}}
 {\hbox{\cmss Z\kern-.4em Z}}{\lower.9pt\hbox{\cmsss Z\kern-.36em Z}}
 {\lower1.2pt\hbox{\cmsss Z\kern-.36em Z}}\else{\cmss Z\kern-.4em
 Z}\fi}

\def\inbar{\,\vrule height1.5ex width.4pt depth0pt}
\def\IC{{\relax\hbox{$\inbar\kern-.3em{\rm C}$}}}
\def\IP{{\relax{\rm I\kern-.18em P}}}
\def\IF{{\relax{\rm I\kern-.18em F}}}
\def\IZ{\relax\ifmmode\mathchoice
{\hbox{\cmss Z\kern-.4em Z}}{\hbox{\cmss Z\kern-.4em Z}}
{\lower.9pt\hbox{\cmsss Z\kern-.4em Z}}
{\lower1.2pt\hbox{\cmsss Z\kern-.4em Z}}\else{\cmss Z\kern-.4em
Z}\fi}
\def\IR{{\relax{\rm I\kern-.18em R}}}
\def\IT{{\mathchoice {\setbox0=\hbox{$\displaystyle\rm
T$}\hbox{\hbox to0pt{\kern0.3\wd0\vrule height0.9\ht0\hss}\box0}}
{\setbox0=\hbox{$\textstyle\rm T$}\hbox{\hbox
to0pt{\kern0.3\wd0\vrule height0.9\ht0\hss}\box0}}
{\setbox0=\hbox{$\scriptstyle\rm T$}\hbox{\hbox
to0pt{\kern0.3\wd0\vrule height0.9\ht0\hss}\box0}}
{\setbox0=\hbox{$\scriptscriptstyle\rm T$}\hbox{\hbox
to0pt{\kern0.3\wd0\vrule height0.9\ht0\hss}\box0}}}}
\def\bbbti{{\mathchoice {\setbox0=\hbox{$\displaystyle\rm
T$}\hbox{\hbox to0pt{\kern0.3\wd0\vrule height0.9\ht0\hss}\box0}}
{\setbox0=\hbox{$\textstyle\rm T$}\hbox{\hbox
to0pt{\kern0.3\wd0\vrule height0.9\ht0\hss}\box0}}
{\setbox0=\hbox{$\scriptstyle\rm T$}\hbox{\hbox
to0pt{\kern0.3\wd0\vrule height0.9\ht0\hss}\box0}}
{\setbox0=\hbox{$\scriptscriptstyle\rm T$}\hbox{\hbox
to0pt{\kern0.3\wd0\vrule height0.9\ht0\hss}\box0}}}}
\def\K{{\cal{K}}}

\def\H{{\cal H}}

\def\tPsi{{\widetilde\Psi}}
\def\OCF{{\overline {\cal F}}}

\def\op{{\overline P}}
\def\oq{{\overline Q}} 

\def\oe{{\overline E}}
\def\talpha{{\widetilde \alpha}}
\def\tbeta{{\widetilde \beta}}
\def\tgamma{{\widetilde \gamma}}
\def\ta{\widetilde\alpha}
\def\tb{\widetilde\beta}
\def\tg{\widetilde\gamma}

\def\rra#1{
  \setbox1=\hbox{\kern10pt${#1}$\kern10pt}
  \,\vbox{\offinterlineskip\hbox to\wd1{\hfil\copy1\hfil}
    \kern 3pt\hbox to\wd1{\rightarrowfill}}\,}

\nref\vgeemen{A. Albano and S. Katz, ``Van Geemen's Families of Lines on Special 
Quintic Threefolds'', Manuscripta Math. {\bf 70} (1991) 183.} 
\nref\AK{A. Albano and S. Katz, ``Lines on The Fermat Quintic Threefold'', 
Trans. AMS. {\bf 324} (1991) 353.}
\nref\ADD{S.K. Ashok, E. Dell'Aquila and D.-E. Diaconescu, 
``Fractional Branes in Landau-Ginzburg Orbifolds'', hep-th/0401135.}
\nref\AL{P. S. Aspinwall and A. E. Lawrence, ``Derived categories and zero-brane stability,'' 
JHEP {\bf 0108}, 004 (2001), hep-th/0104147.}
\nref\B{A.A. Beilinson, ``Coherent Sheaves on $\IP^n$ and Problems in Linear 
Algebra'', Funk. An. {\bf 12} (1978) 68.} 
\nref\Bo{A. I. Bondal, ``Helixes, Representations of Quivers and Koszul Algebras'', 
{\it Helices and Vector Bundles}, A.N. Rudakov ed, London Mathematical Society 
Lecture Series {\bf 148} 1990.}
\nref\BDLR{I. Brunner, M. R. Douglas, A. E. Lawrence and C. Romelsberger,
``D-branes on the quintic,'' JHEP {\bf 0008}, 015 (2000), hep-th/9906200.}
\nref\BHLS{I. Brunner, M. Herbst, W. Lerche and B. Scheuner, ``Landau-Ginzburg
Realization of Open String TFT'', hep-th/0305133.}
\nref\BSi{I. Brunner and V. Schomerus,
``D-branes at Singular Curves of Calabi-Yau Compactifications'', 
JHEP {\bf 04} (2000) 020, hep-th/0001132.}
\nref\BSii{I. Brunner and V. Schomerus, 
``On Superpotentials for D-Branes in Gepner Models'', JHEP {\bf 10} (2000) 016,hep-th/0008194}
\nref\obsII{H. Clemens, ``Cohomology and Obstructions II: Curves on K-Trivial Threefolds'', 
math.AG/0206219.} 
\nref\CK{H. Clemens and H.P. Kley, ``Counting curves which move with threefolds'', 
J. Alg. Geom. {\bf 9} (2000) 175, math.AG/9807109.}
\nref\DED{D.-E. Diaconescu, ``Enhanced D-brane Categories from String Field Theory'', 
JHEP {\bf 06} (2001) 016, hep-th/0104200.}
\nref\DT{S.K. Donaldson and R.P. Thomas, ``Gauge Theories in Higher Dimensions'', 
{\it The Geometric Universe,} 31, Oxford Univ. Press, Oxford 1998.}
\nref\D1{M. R. Douglas, ``D-branes, categories and N = 1 supersymmetry,'' 
J.\ Math.\ Phys.\  {\bf 42}, 2818 (2001), hep-th/0011017.}
\nref\DGJT{M. R. Douglas, S. Govindarajan, T. Jayaraman and A. Tomasiello, 
``D-branes on Calabi-Yau manifolds and superpotentials,'', hep-th/0203173.}
\nref\E{D. Eisenbud, {\it Commutative Algebra with a View Toward Algebraic Geometry,} 
Springer-Verlag, New York 1995.} 
\nref\FukI{K. Fukaya, `` Deformation theory, homological algebra and mirror symmetry,''
{\it Geometry and Physics of Branes,} 121-209, Ser. High Energy Phys. Cosmol. Gravit.,
IOP Bristol 2003.}
\nref\FukII{K. Fukaya, Y.-G. Oh, H. Ohta and K. Ono, ``Lagrangian Intersection 
Floer Theory -- Anomaly and Obstruction -- ``, preprint.} 
\nref\GZ{M.R. Gaberdiel and B. Zwiebach, ``Tensor constructions of open string theories. I. Foundations,''
Nucl. Phys. {\bf B505} (1997) 569, hep-th/9705038.}
\nref\GH{P.Griffiths and J. Harris, {\it Principles of Algebraic Geometry,} Wiley-Interscience, 
1994.}
\nref\JH{J. Harris, {\it Algebraic Geometry,} Springer-Verlag, New York, 1992.}
\nref\RH{R. Hartshorne, {\it Algebraic Geometry,} Springer-Verlag, New York, 1977.}
\nref\HLL{M. Herbst, C.I. Lazaroiu and W. Lerche, ``Superpotentials, $A_\infty$ Relations 
and WDVV Equations for Open Topological Strings'', hep-th/0402110.}
\nref\H{K. Hori, ``Linear Models of Supersymmetric D-Branes'', hep-th/0012179.}
\nref\IV{K. Intriligator and C. Vafa, ``Landau-Ginzburg Orbifolds'', Nucl. Phys. {|bf B339} 
(1990) 95.} 
\nref\KLI{A. Kapustin and Y. Li, ``D-Branes in Landau-Ginzburg Models and Algebraic Geometry'', 
hep-th/0210296.}
\nref\KLII{A. Kapustin and Y. Li, 
``Topological Correlators in Landau-Ginzburg Models with Boundaries'', 
hep-th/035136.}
\nref\KLIII{A. Kapustin and Y. Li, 
``D-Branes in Topological Minimal Models: The Landau-Ginzburg Approach'', 
hep-th/0306001.}
\nref\ESii{S. Katz and E. Sharpe, ``D-branes, open string vertex operators, and Ext groups'', 
Adv. Theor. Math. Phys. {\bf 6} (2003) 979, hep-th/0208104}%
\nref\ESiii{S. Katz, T. Pantev and E. Sharpe, ``D-branes, orbifolds, and Ext groups'', 
Nucl. Phys. {\bf B673} (2003) 263, hep-th/0212218}%
\nref\KKLMi{S. Kachru, S. Katz, A.E. Lawrence and J. McGreevy, 
``Open String Instantons and Superpotentials'', 
Phys. Rev. {\bf D62} (2000) 026001, hep-th/9912151.}
\nref\KKLMii{S. Kachru, S. Katz, A.E. Lawrence and J. McGreevy, 
`` Mirror Symmetry for Open Strings'', 
Phys. Rev. {\bf D62} (2000) 126005, hep-th/0006047.} 
\nref\versal{S. Katz, ``Versal Deformations and Superpotentials for Rational Curves in Smooth 
Threefolds'', math.AG/0010289.} 
\nref\BK{B. Keller, ``Introduction to $A$-infinity algebras and modules,'' 
Homology Homotopy Appl. {\bf 3} (2001), 1, math.RA/9910179.}
\nref\curves{H.P. Kley, ``On The Existence of Curves in K-Trivial Threefolds'', 
math.AG/9811099.}
\nref\rigid{H.P. Kley, ``Rigid curves in complete intersection Calabi-Yau threefolds'', 
Comp. Math. {\bf 123} (2000) 185.} 
\nref\K{M. Kontsevich, ``Homological Algebra of Mirror Symmetry'', 
alg-geom/9411018.}
\nref\CLi{C.I. Lazaroiu, ``Unitarity, D-brane dynamics and D-brane categories'', 
JHEP {\bf 12} (2001) 031, hep-th/0102183.}
\nref\CLii{C.I. Lazaroiu, ``Generalized complexes and string field theory'', 
JHEP {\bf 06} (2001) 052, hep-th/0102122.}
\nref\CLiii{C.I. Lazaroiu, ``String field theory and brane superpotentials'', 
JHEP {\bf 10} (2001) 018, hep-th/0107162.}
\nref\CLiv{C.I. Lazaroiu, ``D-Brane Categories'', Int. J. Mod. Phys. {\bf A18} (2003) 
5299, hep-th/0305095.}
\nref\LR{C.I. Lazaroiu and R. Roiban, ``Holomorphic Potentials for Graded D-Branes'', 
JHEP {\bf 0202} (2002) 038, hep-th/0110288.}
\nref\SAM{S.A. Merkulov, ``Strong Homotopy Algebras of a K\"ahler Manifold'', 
Internat. Math. Res. Notices {\bf 3} (1999) 153.} 
\nref\O{D. Orlov, ``Triangulated Categories of Singularities and 
D-Branes in Landau-Ginzburg Orbifolds'', math.AG/0302304.}
\nref\Poli{A. Polishchuk, ``Homological mirror symmetry with higher products,'' 
{\it Winter School on Mirror Symmetry, Vector Bundles and Lagrangian Submanifolds (Cambridge, MA, 1999),}
247-259, Amer. Math. Soc., Providence, RI, 2001, math.AG/9901025.}
\nref\Polii{A. Polishchuk, ``$A_{\infty}$-Structures on an Elliptic Curve'', math.AG./0001048.}
\nref\ESi{E. Sharpe, ``D-Branes, Derived Categories, and Grothendieck Groups'', 
Nucl. Phys. {\bf B561} (1999) 433, hep-th/9902116.}%
\nref\EWii{E.
Witten, ``Chern-Simons Gauge Theory as a String Theory'',
``The Floer Memorial Volume'', H. Hofer, C.H. Taubes, A. Weinstein
and E. Zehnder, eds, Birkh\"auser 1995, 637,
hep-th/9207094.}
\nref\W1{E. Witten, ``Phases of N = 2 theories in two dimensions,'' 
Nucl.\ Phys.\ B {\bf 403}, 159 (1993), 
hep-th/9301042.}
\nref\EWiii{E. Witten, ``Branes And The Dynamics Of QCD'', Nucl. Phys. {\bf B507} (1997) 658, 
hep-th/9706109.}
\Title{
\vbox{
\baselineskip12pt
\hbox{hep-th/0404167}}}
{\vbox{\vskip 37pt
\vbox{\centerline{Obstructed D-Branes in Landau-Ginzburg Orbifolds}}}}
\vskip 15pt
\centerline{Sujay K. Ashok, Eleonora Dell'Aquila,  
Duiliu-Emanuel Diaconescu and Bogdan Florea\footnote{$^{}$}
{{{\tt ashok,dellaqui,duiliu,florea@physics.rutgers.edu}}}}
\bigskip
\medskip
\centerline{{\it Department of Physics and Astronomy,
Rutgers University,}}
\centerline{\it Piscataway, NJ 08855-0849, USA}
\bigskip
\bigskip
\bigskip
\bigskip
\smallskip
\noindent 
We study deformations of Landau-Ginzburg D-branes corresponding to obstructed rational 
curves on Calabi-Yau threefolds. We determine D-brane moduli spaces and D-brane 
superpotentials by evaluating higher products up to homotopy in the Landau-Ginzburg 
orbifold category. For concreteness we work out the details for lines on a perturbed 
Fermat quintic. In this case we show that our results reproduce the local analytic 
structure of the Hilbert scheme of curves on the threefold.  

\vfill
\Date{April 2004}

\newsec{Introduction} 

D-branes wrapping holomorphic cycles in Calabi-Yau threefolds often give 
rise to interesting $N=1$ field theories. The low energy interactions 
are specified by an effective superpotential on the space of massless fields. 
In principle, the effective tree level superpotential is determined by 
topological disc correlators with an arbitrary number of boundary insertions. Its 
dependence on closed string moduli is captured by topological disc correlators with bulk 
and boundary insertions. Such correlators are typically very hard to evaluate by direct 
computation, so one is often compelled to search for alternative methods.  

According  to \refs{\D,\AL,\ESi}, holomorphic branes on Calabi-Yau manifolds 
should be properly regarded as derived objects. In this context, 
the tree level superpotential is encoded in the deformation 
theory of objects in the derived category, which is in turn determined by higher 
$A_\infty$ products \refs{\SAM,\Poli,\Polii,\FukI}.
This was explained in the physics literature in \CLiii\ and 
recently derived from the conformal field theory point of view in \HLL.\ 
Physically, this means one has to compute the tree level effective 
action of holomorphic Chern-Simons theory on a Calabi-Yau manifold. This computation is 
practically untractable for compact Calabi-Yau manifolds because the gauge fixing 
procedure relies on the choice of a metric. 

There are several alternative approaches to this problem. One can attempt a direct 
CFT computation of boundary correlators \refs{\BDLR,\BSii}, but this method has given 
rather restricted results so far. A more powerful approach has been recently developed 
in \HLL.\ There it has been shown that topological disc correlators with bulk and boundary 
insertions are subject to a series of algebraic constraints which can fix them completely  
at least for {\bf B}-branes in minimal models. We will not discuss this in detail here, 
although it would be interesting to apply this method to D-branes on Calabi-Yau manifolds. 
A different construction has appeared in the mathematics literature \refs{\obsII,\versal}
where the Hilbert scheme of curves on a Calabi-Yau threefold 
is locally described as the critical locus of an analytic function.
Moreover, it was 
shown in \obsII\ that the analytic function in question is essentially the three-chain integral of 
\DT\ known to physicists as the membrane superpotential in 
MQCD \refs{\EWiii}. However this function is very hard to calculate for curves on compact Calabi-Yau 
manifolds, even as a perturbative expansion. More conceptually, the relation between this superpotential  
and perturbative open string computations is rather obscure at the present stage. 

The main point of the present paper is to develop a new approach to D-brane deformations and 
superpotentials based on open string Calabi-Yau/Landau-Ginzburg correspondence. 
{\bf B}-branes in Landau-Ginzburg orbifolds can be described in terms of matrix factorizations 
of the Landau-Ginzburg superpotential \refs{\KLI,\O,\BHLS}.  
These objects form a triangulated category in which one can perform explicit 
computations for the higher products by pure algebraic manipulations. In order to 
illustrate this principle, we determine the higher products for Landau-Ginzburg 
D-branes corresponding to obstructed curves on Calabi-Yau threefolds. 
However, this approach can be very well implemented in more general situations. 
We also show that the dependence of higher products on complex structure moduli is 
encoded in a simple deformation of the $A_\infty$ structure inferred from the  
geometric context. This data determines a D-brane superpotential up to homotopy.
For concreteness, the computations will be carried out for obstructed lines on the 
Fermat quintic threefold. As a consistency test of our results, we check that the 
resulting D-brane moduli space is isomorphic (as a germ of analytic space) with 
the Hilbert scheme of lines on the quintic. 

The paper is structured as follows. In section $2$ we review some aspects of 
Landau-Ginzburg/Calabi-Yau correspondence for branes, and present the Landau-Ginzburg 
realization of lines on Calabi-Yau threefolds. Section $3$ consists of a general 
discussion of higher products in the Calabi-Yau as well as Landau-Ginzburg 
phase. We explain the general principles of our approach and outline a computational 
algorithm. The dependence on complex 
structure moduli is the main theme of section $4$. 
In section $5$ we give a detailed treatment for lines on a perturbed Fermat 
quintic as an illustration of the method. 

{\it Acknowledgments} We would like to thank Mike Douglas for very helpful discussions. 
The work of D.-E.D. is partially supported by DOE grant DE-FG02-96ER40949 
and an Alfred P. Sloan foundation fellowship. The work of B.F. is supported by 
DOE grant  DE-FG02-96ER40949.

\newsec{Landau-Ginzburg D-Branes and Curves on Calabi-Yau Threefolds} 

The goal of this section is to find the Landau-Ginzburg description of D-branes wrapping 
rational curves in Calabi-Yau threefolds. We will review the Calabi-Yau/Landau-Ginzburg 
correspondence for D-brane categories and explain the relevant constructions.  

Since we are interested in 
Calabi-Yau manifolds which admit a pure Landau-Ginzburg phase, we will restrict our 
considerations to 
hypersurfaces in weighted projective spaces. Let $X$ be such a hypersurface in a weighted 
projective space $WP^{w_0, \ldots , w_n}$ determined by a quasihomogeneous polynomial
$W_{LG}(x_0,\ldots, x_n)$ of degree $d=w_0+\ldots+ w_n$. Using the linear sigma model 
realization, 
one can deform the nonlinear sigma model on $X$ to a Landau-Ginzburg orbifold with 
superpotential 
$W_{LG}$. The orbifold group is $G=\IZ/d$ acting on the chiral fields as 
$x_i\ra \omega^{w_i} x_i$ for $i=0,\ldots,n$, where $\omega$ is $d$-th root of unity. 
For the purpose of computing the superpotential, it suffices to consider the 
{\bf B}-twisted topological model. 

In the geometric phase the category of ${\bf B}$-type topological branes is 
the derived category $D^b(X)$. In the Landau-Ginzburg phase, the D-brane category can be given 
a very elegant algebraic description in terms of matrix factorizations of $W_{LG}$ 
\refs{\KLI,\O,\BHLS}. 
Standard decoupling arguments suggest that the two categories should be physically equivalent, but 
a rigorous mathematical result along these lines has not appeared so far in the literature.
Note that there must be a subtlety in formulating such an equivalence of categories since 
the Landau-Ginzburg category is $\IZ/2$ graded while the derived category is $\IZ$ 
graded. Nevertheless, 
a physical correspondence between certain classes of 
Landau-Ginzburg and geometric branes has been found in  
\refs{\ADD}. We will review the relevant points below. 

Matrix factorizations of $W_{LG}$ are pairs of finitely generated projective modules $P_1,P_0$ over 
the polynomial ring $R=\IC[x_0,\ldots,x_n]$ equipped with two $R$-module homomorphisms 
$\xymatrix{P_1 \ar@<1ex>[r]^{p_1}& P_0\ar@<1ex>[l]^{p_0}\\}$ so that $p_1p_0=p_0p_1=W_{LG}$. 
In the present paper is suffices to take $P_1,P_0$ to be free $R$-modules. It has been shown 
in \refs{\KLI,\O} that matrix factorizations form a $\IZ/2$ graded triangulated category 
$\CC_{W_{LG}}$.  
For future reference, let us recall the construction of morphisms in $\CC_{W_{LG}}$. 
Given two objects 
$\op, \oq$, one forms a $\IZ/2$ graded cochain complex $(\IH(\op,\oq),D)$ where 
\eqn\complexA{ 
\IH(\op,\oq)=\hbox{Hom}(P_1\oplus P_0,  Q_1\oplus Q_0)
=\bigoplus_{i,j=0,1} \hbox{Hom}(P_i,Q_j).} 
The grading is given by $(i-j)$ mod 2 and the differential $D$ is determined by its action on 
degree $k$ homogeneous elements 
\eqn\complexB{ 
D\cdot\Phi = q\cdot\Phi -(-1)^k \Phi\cdot p}
where $p=p_1\oplus p_0:P_1\oplus P_0\ra P_1\oplus P_0$, $q=q_1\oplus q_0 : Q_1\oplus Q_0 \ra 
Q_1\oplus Q_0$.
This data defines a DG-category $\CP_{W_{LG}}$ \refs{\KLI,\O}. 
The D-brane category $\CC_{W_{LG}}$ is the 
category associated to $\CP_{W_{LG}}$ by taking the space of morphisms between two 
objects $(\op,\oq)$ to 
be the degree zero cohomology $H^0(\IH(\op,\oq))$ of the complex \complexA.\
We will use the shorthand notation $H^k(\op,\oq)$, $k=0,1$ for the cohomology groups.

For Landau-Ginzburg orbifolds, one should employ a $G$-equivariant version of this 
construction. The modules $P_1,P_0$ should be endowed with representations $R_1,R_0$ of the orbifold group $G$ so that $p_0, p_1$ are $G$-equivariant $R$-module homomorphisms.
Then there is an induced $G$-action on the cochain complex \complexA\ so that $D$ is 
$G$-equivariant. The space of morphisms in the orbifold  category 
$\CC_{W,\rho}$ is given by the $G$-fixed part of the cohomology groups $H^0(\op,\oq)$.

One of the main problems of this approach is that matrix factorizations are quite hard to 
construct in practice. A very efficient tool employed in \ADD\ is the tensor product of two 
factorizations. Namely, suppose 
$$\op = \left(\xymatrix{P_1 \ar@<1ex>[r]^{p_1}& P_0\ar@<1ex>[l]^{p_0}\\}\right), \qquad 
\oq=\left(\xymatrix{Q_1 \ar@<1ex>[r]^{q_1}& Q_0\ar@<1ex>[l]^{q_0}\\}\right)$$ 
are two factorizations 
associated to LG polynomials $U(x_i),V(x_i)$, $i=1,\ldots, n$. 
Then one can construct a matrix factorization $\op\otimes \oq$ for 
$W_{LG}(x_i)=U(x_i)+V(x_i)$
by taking a $\IZ/2$ graded tensor product of $\op, \oq$. More precisely, 
\eqn\tensorA{ \eqalign{ 
& (\op\otimes \oq)_1 = P_1 \otimes_{R} Q_0 \oplus 
P_0 \otimes_{R} Q_1\cr
& (\op\otimes \oq)_0 = P_0\otimes_{R} Q_0 \oplus 
P_1 \otimes_{R} Q_1.\cr}}
The maps 
$\xymatrix{(\op\otimes \oq)_1 \ar@<1ex>[r]^{r_1}&(\op\otimes \oq)_0 \ar@<1ex>[l]^{r_0}\\}$ are given by 
\eqn\tensorB{ 
r_1=\left[\matrix{ p_1\otimes \II & \II\otimes q_1 \cr \II\otimes q_0 & -p_0\otimes 
\II\cr}\right]
\qquad 
r_0=\left[\matrix{ p_0\otimes \II & \II\otimes q_1 \cr \II\otimes q_0 & -p_1\otimes \II\cr}
\right].}
The tensor product can be naturally extended to equivariant objects, as explained in \ADD.\ 

This construction is especially effective for LG superpotentials $W$ in Fermat form
\eqn\fermat{
W_{LG}(x_0,\ldots, x_n) = x_0^{d_0} + \ldots + x_n^{d_n}}
where $w_id_i =d$ for each $i=0,\ldots,n$. 
In this case one can construct matrix factorizations of $W_{LG}$ by taking tensor products of one or two 
variable building blocks which can be described as follows. 

Consider a one variable polynomial $W_{LG}(x)=X^{d'}$ with $d'=d/w$ for some weight $w$. 
The category $\CC_W$ is generated by rank one 
objects of the form 
\eqn\blocksA{ 
\om_{l} =  \left(\xymatrix{\IC[x] \ar@<1ex>[r]^{x^l}& \IC[x]\ar@<1ex>[l]^{x^{d'-l}}\\}\right).}
In the orbifold theory we have to specify irreducible representations $R_1, R_0$ 
of $G=\IZ/d$ on the free modules $M_1=M_0=\IC[x]$
so that the maps are equivariant. If $R_1$ is multiplication by $\omega^\alpha$, 
$\alpha \in \{0,\ldots, d-1\}$, it follows that $R_0$ must be multiplication 
by $\beta = \alpha + l$. Therefore it suffices to specify one integer $\alpha$ mod $d$. 
The resulting objects will be denoted by $\om_{l,\alpha}$. 

For a two variable polynomial $W_{LG}(x,y)=x^{d'}+y^{d'}$, we can construct similar objects 
\eqn\blocksB{ 
\op_{\eta} = \left(\xymatrix{\IC[x,y] \ar@<1ex>[r]^{x-\eta y}& 
\IC[x,y]\ar@<1ex>[l]^{Q_\eta(x,y)}\\}\right)}
where $\eta$ is a $d'$-th root of $-1$ and $Q_\eta(x,y) = (x^{d'}+y^{d'})/(x-\eta y)$. 
In the orbifold theory objects will be labeled by an additional integer $\alpha$ mod $d$ 
specifying the action of $G$ on $P_1=\IC[x,y]$. 

Given a Landau-Ginzburg superpotential of the form \fermat\ one can construct matrix 
factorizations by writing it as a sum of one and two variable polynomials and 
taking tensor products \ADD.\ 
The resulting objects are very interesting from a geometric point of view. 
In particular one can take the tensor product of one variable building blocks 
$\om_{1,\alpha_i}^{(i)}$ associated to the monomials $x_i^{d_i}$ in $W$. It turns out 
that two such objects $\otimes_{i=0}^n \om_{1,\alpha_i}^{(i)}$, 
$\otimes_{i=0}^n \om_{1,\beta_i}^{(i)}$ 
are isomorphic if $\sum_{i=0}^n \alpha_i = \sum_{i=0}^n \beta_i\ \hbox{mod}\ d$. 
Therefore we obtain a collection of Landau-Ginzburg branes $\oe_\alpha$ 
labeled by a single integer $\alpha = \sum_{i=0}^n \alpha_i$ mod $d$. These objects have 
been identified with the Gepner model fractional branes in \ADD.\ Analytic continuation 
to the geometric phase relates the $\oe_\alpha$ to a collection of sheaves $E_\alpha$ 
on $X$ obtained by restriction of an exceptional collection on the ambient space 
$WP^{w_0,\ldots, 
w_n}$. For example if $X$ is the Fermat quintic in $\IP^4$, $E_\alpha = \Omega^\alpha(\alpha)$. 

Taking tensor products of some combination of one and two variable building blocks results 
in objects with different geometric interpretation. The pattern emerging from the examples 
studied in \ADD\ is the following. 
Pick an arbitrary subset $I\subset \{0,\ldots, n\}$ so that the complement $I^\circ$ 
has an even number of elements. Moreover, pick some arbitrary decomposition of $I^\circ$ 
into pairs $\{i,j\}$ with $d_i=d_j$ so that no two pairs share a common element and the 
union of all 
$\{i,j\}$ is $I^\circ$. Denote the set of all such pairs by $P$. 
Then we can take the tensor product 
\eqn\blocksC{
\bigotimes_{i\in I} \om_{1,\alpha_i}^{(i)} \otimes \bigotimes_{(i,j)\in P} 
\op_{\eta_{ij}, \alpha_{ij}}^{(i,j)}.}
It is easy to check that such objects are again classified by a single integer $\alpha = 
\sum_{i\in I} \alpha_i + \sum_{(i,j)\in P} \alpha_{ij}\ \hbox{mod}\ d$. 
These correspond to a collection of derived objects $F_\alpha$ in the geometric 
phase which form an orbit of the Landau-Ginzburg monodromy transformation acting on 
the derived category $D^b(X)$. 

Given the examples studied in \ADD\ we conjecture that 
the $F_\alpha$ can be constructed as follows.  
Let $F$ be the subvariety of $X$ determined by the equations 
\eqn\blocksD{ 
x_i=0,\ i\in I,\qquad x_i-\eta_{ij}x_j =0,\ (i,j)\in P}
in the ambient weighted projective space, where $\eta_{ij}$ is a $d_i$-th root of $-1$. 
The structure sheaf of $F$ determines an object ${\underline \CO_F}$ in $D^b(X)$ supported in 
degree zero. We conjecture that the $F_\alpha$ can be obtained (up to an overall shift) 
by $d-1$ successive applications 
of the Landau-Ginzburg monodromy transformation 
${\bf M}_{LG}$ to the object ${\underline \CO_F}$. 

For a better conceptual formulation of this conjecture, one can use the following result of \O.\ Suppose we are given a Landau-Ginzburg superpotential 
$W_{LG}:\IC^{n+1}\ra C$ with an isolated critical 
point at the origin. Let $S_0$ denote the fiber of $W_{LG}$ over $0\in \IC$. Then  
the main statement of \O\ is that the D-brane category 
$\CC_{W_{LG}}$ is equivalent to the so-called category of the singularity 
$D_{Sg}(S_0)$. $D_{Sg}(S_0)$ is constructed by taking the quotient of the 
bounded derived category $D^b(S_0)$ by the full subcategory of perfect complexes. 
A perfect complex is a finite complex of locally free sheaves. If $S_0$ were nonsingular, the quotient would be empty, since in that case any object in $D^b(S_0)$ would have a
finite locally free resolution. Therefore $D_{Sg}(S_0)$ depends only on the singular points
of $S_0$. The equivalence functor $\CC_{W_{LG}} \ra D_{Sg}(S_0)$ associates to an object 
$\Big(\xymatrix{P_1 \ar@<1ex>[r]^{p_1}& P_0\ar@<1ex>[l]^{p_0}\\}\Big)$
the one term complex defined by the cokernel of $p_1$ regarded as a coherent 
$R/W_{LG}$-module.
For factorizations of the form \blocksC\ one can show 
that $\hbox{Coker}(p_1)$ is isomorphic to 
the quotient ring $R/(x_i,x_i-\eta_{ij}x_j)_{i\in I, (i,j)\in P}$. 
The proof is very similar to that performed in section six of \ADD\ for $D0$-brane 
factorizations. 
Therefore we are lead to a direct relation 
between the factorization $\of$ and the structure ring of the associated geometric object. 
This conjecture should hold for any Calabi-Yau hypersurface in a weighted projective space, 
but we do not know a general proof. 
In principle, it should follow from a rigorous mathematical relation between 
the Landau-Ginzburg orbifold category and $D^b(X)$ if such a relation were explicitly 
known. For the purpose of this paper, we will be content to check it for specific examples. 

Note that we can formulate a more general conjecture for complete intersections $F$ of the form 
\eqn\complintA{
f_a(x_i) =0,\qquad a=1,\ldots, A}
in $WP^{w_0,\ldots,w_n}$ which lie on $X$. Assuming that $F$ is irreducible, it follows from
Hilbert Nullstellensatz that we must have a decomposition 
\eqn\complintB{
W_{LG} = \sum_{a=1}^A q_af_a}
for some quasi-homogeneous polynomials $q_a(x_i)$. Then we conjecture that the Landau-Ginzburg 
monodromy orbit associated to $F$ is described at the Landau-Ginzburg point by 
factorizations of the form 
\eqn\complintC{
\of=\bigotimes_{a=1}^A \left(\xymatrix{R \ar@<1ex>[r]^{f_a}& R\ar@<1ex>[l]^{q_a}\\}\right).}
This construction was used in \ADD\ in order to construct deformations of the $D0$-brane in the Landau-Ginzburg category.

A check of this conjecture for a point $F=\{x_0-\eta_{01} x_1=0,x_2=x_3=x_4=0\}$
on the Fermat quintic was performed in \ADD.\ The argument assumes that the geometric 
interpretation of the fractional branes ${\oe}_\alpha$ are known and is based 
on the Beilinson correspondence. More precisely, one can show that there is an isomorphism between the endomorphism algebra 
$${\overline \CA} = \hbox{End}\left(\bigoplus_{\alpha=0}^{d-1}{\oe}_\alpha [\alpha]\right)$$
in the Landau-Ginzburg phase 
and the algebra 
$$ \CA = \hbox{End}\left(\bigoplus_{\alpha=0}^{d-1}E_\alpha \right)$$
determined by the exceptional collection. 
Beilinson's theorem implies that any derived object $\CF$ in $D^b(\IP^4)$ 
is uniquely determined up 
to isomorphism by the left $\CA$-module structure of the graded vector space 
\eqn\RHomA{
{\bf R}\hbox{Hom}\left(\bigoplus_{\alpha=0}^{d-1}E_\alpha, \CF\right) = 
\oplus_{k\in \IZ} \hbox{Hom}\left(\bigoplus_{\alpha=0}^{d-1}E_\alpha, \CF[k]\right)[k].} 
This allows us to determine the derived object corresponding to any Landau-Ginzburg D-brane ${\overline F}$ knowing the left ${\overline \CA}$-module structure of the $\IZ/2$ graded vector space 
\eqn\RHomB{ 
{\bf R}\hbox{Hom}\left(\bigoplus_{\alpha=0}^{d-1}{\overline E}_\alpha[\alpha], 
{\overline F}\right) 
= \bigoplus_{k\in \IZ/2} \hbox{Hom}\left(\bigoplus_{\alpha=0}^{d-1}\oe_\alpha[\alpha], 
{\overline F}[k]\right)[k].}

Employing the same argument, one can check the above conjecture for other objects \blocksC\ in concrete examples. For example, one can construct the $\IZ/5$ orbit of a line  
\hbox{$F=\{x_0-\eta_{01}x_1=x_2-\eta_{23} x_3 = x_4 =0\}$} 
on the Fermat quintic by taking a tensor product
\eqn\blocksE{ 
\op^{(0,1)}_{\eta_{01},\alpha_{01}}\otimes \op^{(2,3)}_{\eta_{23},\alpha_{23}} 
\otimes \om^{(4)}_{1,\alpha_4}.}
The proof is very similar to the D0-brane case considered in section 5 of \ADD,\ hence we will omit the details. 

To conclude this section, we will add some remarks concerning the behavior of morphisms between branes under the Calabi-Yau/Landau-Ginzburg correspondence. Given two factorizations $\of,\of'$ corresponding to two objects $F,F'$ in $D^b(X)$, one would like to know 
the relation between the cohomology groups $H^k(\of,\of')$, $k=0,1$ and the morphism spaces $\hbox{Hom}_{D^b(X)}(F,F'[l])$, $l\in \IZ$. For all examples described in this section one can check that 
\eqn\morphA{\eqalign{
H^0(\of,\of')\simeq \hbox{Hom}_{D^b(X)}(F,F')\oplus \hbox{Hom}_{D^b(X)}(F,F'[2])\cr
H^1(\of,\of')\simeq \hbox{Hom}_{D^b(X)}(F,F'[1])\oplus \hbox{Hom}_{D^b(X)}(F,F'[3]).\cr}}
These relations reflect the difference in grading between the two categories. 

\newsec{D-Branes, Deformations and Higher Products}

In this section we develop a computational approach to D-brane deformation theory 
in the Landau-Ginzburg orbifold category. We start with a review of deformation theory 
and higher products in the derived category associated to the geometric phase. 
Then we set up the problem and explain the main idea of our construction in the Landau-Ginzburg 
phase. 

Let $F$ be a line on a Calabi-Yau hypersurface $X$ in a weighted projective space. We will assume that $X$ is smooth, at least near $F$. (If not, one may have to resolve the 
singularities of $X$ induced by the orbifold singularities of the ambient space.) 
Physically we would like to think of $F$ as the world-volume of a ${\bf B}$-twisted brane. 
This means that we have a boundary {\bf B}-type topological field theory (TFT) 
associated to the pair $(X,F)$, 
which can be described from two different (but equivalent) points of view. 

From the point of view of \refs{\D,\AL}, a boundary ${\bf B}$-topological model 
should be thought 
of as a derived object in $D^b(X)$. In our case, this object is the one 
term complex $\underline
{\CO_F}$ supported in degree zero. Moreover, since isomorphic derived objects 
define isomorphic 
boundary TFT's, we can replace $\underline{\CO_F}$ by a locally free resolution 
${\CF}^{\bullet}$. 
In this formulation, the boundary chiral ring is isomorphic to the Ext algebra 
\eqn\bdryAA{\bigoplus_{k=0}^3\hbox{Ext}^k(\CO_F, \CO_F).}
In particular there is a grading on physical states that assigns ghost number $k$ to elements in $\hbox{Ext}^{k}(\CO_F, \CO_F)$. 

The second point of view relies on the sigma model description of the boundary 
TFT following \refs{\EWii}. In that case, one has to specify appropriate boundary conditions 
for the sigma model and compute the massless spectrum with standard boundary 
TFT methods \ESii.\  
In this approach one finds the same spectrum of physical operators \bdryAA\ realized as the 
limit of a local to global spectral sequence with second term \ESii\ 
\eqn\bdryA{ 
E_2^{p,q} = H^p(F,\Lambda^q(N_{F/X})).}
This proves the equivalence of these two points of view. 
If $F$ is a curve, all higher differentials are zero, and we find
\eqn\bdryB{ 
\hbox{Ext}^k(\CO_F, \CO_F)\simeq \bigoplus_{p+q=k} H^p(F,\Lambda^q(N_{F/X})).}

Boundary topological field theories admit boundary marginal perturbations which are classified 
by ghost number one elements in the chiral ring \bdryAA.\ These are infinitesimal 
first order deformations of the 
theory induced by perturbations of the boundary conditions keeping the 
underlying bulk theory fixed. 
It is common knowledge that not all first order infinitesimal perturbations 
can be integrated to finite deformations of the boundary TFT's. Some deformations 
are marginal at first order, but do not remain marginal at higher orders. 
Those deformations which are marginal to all orders are called exactly marginal. 
This phenomenon is 
encoded by a holomorphic superpotential on the space of all massless ghost number one modes. 
The critical locus of the superpotential defines the local moduli space of TFT's as 
an analytic subspace of the linear space of massless modes.
The superpotential is determined in principle by topological disc correlators with an 
arbitrary number of physical boundary insertions. 

A short computation based on \bdryB\ shows that the infinitesimal boundary deformations 
are classified by $H^0(F, N_{F/X})$. Therefore they are in one to one 
correspondence with infinitesimal deformations of the curve $F$ in $X$. Not all such infinitesimal deformations are integrable to all orders. Those 
that fail to be integrable to all orders are called obstructed, while the 
remaining ones are called unobstructed. Using Kuranishi theory, one can construct 
the versal deformation space of the curve $F$ in $X$, which is roughly the space 
spanned by all unobstructed deformations. Technically, the versal moduli space is cut out 
by formal power series in several variables, i.e. it is a germ of analytic space. 
Alternatively, one can think of it as the formal completion of the Hilbert scheme 
of curves on $X$ at the point $F$. We will loosely refer to it as the local D-brane 
moduli space. The equations of the moduli space 
are determined by the $A_\infty$ products
on the endomorphism algebra of the object ${\underline \CO_F}$ in the derived category 
$D^b(X)$ (see for example \FukI\ for a good exposition.)

Given the correspondence between physics and geometry it follows that 
the D-brane superpotential should be determined by the $A_\infty$ structure 
on the derived category. This connection can be made very precise in the framework of 
string field theory \refs{\EWii,\CLiii} which will be reviewed next.  

\subsec{Holomorphic Chern-Simons Theory} 

For simplicity, let us address a similar question for topological {\bf B}-branes 
described by a holomorphic bundle $E$ on $X$. 
The dynamics of off-shell open string modes in this model is captured the 
holomorphic Chern-Simons action \EWii\ which defines a cubic string field theory. 
In order to write down this action, we must 
regard $E$ as a $C^\infty$ vector bundle equipped with a connection $A$ so that 
$F^{0,2}(A)=0$. The off-shell 
boundary fields form an associative algebra $\CV=\oplus_{p=0}^3 \Omega^{0,p}(X,\hbox{End}(E))$, 
where $p$ represents the ghost number. The string field theory action for ghost 
number one states $a\in \Omega^{0,1}(X,\hbox{End}(E))$ takes the form 
\eqn\suppotA{ 
S_{CS}(a) = \int_{X} \Omega_X \hbox{Tr}\left({1\over 2} a{\overline \partial}_A a + {1\over 3} 
a^3\right),}
where ${{\overline \partial}_A}$ denotes the $(0,1)$ part of the covariant derivative 
with respect to 
the background connection $A$ on $E$. From a physical point of view, ${\delbar}_A$ represents the $BRST$ operator $Q$ acting on off-shell states. Since 
$Q$ satisfies the graded Leibniz rule, it defines a structure of differential graded 
(DG) algebra on $\CV$. Note that the massless spectrum of the theory is parameterized by the 
graded vector space 
$H=\oplus_{p=0}^3 H^{0,p}(X,\hbox{End}(E))$, and the boundary chiral ring structure is defined by 
the Yoneda pairing on $H$. The physical on-shell operators in string field  
theory correspond to elements of degree one in $H$, that is cohomology classes in 
$H^{0,1}(X,\hbox{End}(E))$. 

In this context, the superpotential can be thought of as the tree level effective action for 
physical massless modes $\Phi\in H^{0,1}(X,\hbox{End}(E))$ \refs{\EWii,\CLiii}, and can be computed as follows. For perturbative computations, we have to fix a metric on the Calabi-Yau 
threefold $X$ and a hermitian structure on $E$ so that $A$ is a unitary connection. 
Then, applying the Hodge Theorem, we can decompose the space of off-shell states into a direct sum 
\eqn\suppotB{ 
\CV = \hbox{Im}(Q) \oplus \CH \oplus \hbox{Im}(Q^\dagger),}
where $\CH$ denotes the space of harmonic forms which are in one-to-one correspondence with 
BRST cohomology classes in $H$. Any off-shell field $a$ can be accordingly written as 
\eqn\suppotC{ 
a = \Phi + \Phi_m.}
where $\Phi$ can be expanded in a basis of harmonic representatives $\omega_i$ 
\eqn\cohomrep{
\Phi=\sum_{i} \psi_i\,\omega_i\,,
}
and $\Phi_m$ takes values in $\hbox{Im}(Q)\oplus\hbox{Im}(Q^\dagger)$.
We also impose the gauge fixing condition $Q^\dagger \Phi_m =0$ in order to eliminate the gauge degrees of freedom parameterized by $\hbox{Im}(Q)$. After gauge fixing, $\Phi_m$ is an off-shell field in $\hbox{Im}(Q^\dagger)$. We will refer to $\Phi$ as a massless field and to $\Phi_m$ as a massive field. The effective action for the massless modes is obtained by substituting \suppotC\ into \suppotA\ and integrating out the massive modes $\Phi_m$ at tree level. The resulting superpotential is a generating functional for tree level Feynman diagrams with arbitrary combinations of massless fields on the external legs. In the topological string theory, the tree level diagrams can be regarded as disc correlators receiving contributions from degenerate discs mapping to infinitely thin ribbon graphs in the target space $X$ \refs{\EWii}. 

From a mathematical point of view, Chern-Simons tree level diagrams can be expressed in terms of higher $A_\infty$ products $m_n :H^{\otimes n}\ra H$, $n\geq 1$ \refs{\SAM,\CLiii} 
satisfying certain generalized associativity conditions. To recall some background \refs{\BK}, 
note that 
a strong $A_\infty$ structure on a $\IZ$-graded vector space $V=\oplus_{p\in \IZ} V^p$ is defined 
by a sequence of linear maps \hbox{$m_n : V^{\otimes n} \ra V$} of degree $2-n$, $n\geq 1$, satisfying the 
generalized associativity conditions 
\eqn\suppotCA{
\sum (-1)^{r+st}m_u\left(\II^{\otimes r} \otimes m_s \otimes \II^{\otimes t}\right)=0.}
The sum in the left hand side of \suppotCA\ runs over all decompositions 
$n=r+s+t$, and $u=r+t+1$. 
This means for example that $m_1^2=0$, hence $m_1$ defines a differential on $V$. Moreover,
$$m_1m_2=m_2\left(m_1\otimes \II + \II\otimes m_1\right),$$ 
hence $m_1$ satisfies the graded 
Leibniz rule with respect to the multiplication defined by $m_2$. The next equation in 
\suppotCA\ imposes a
generalized associativity condition on $m_2$ and so on. Note that if the formulae \suppotCA\ 
are applied to elements, one has additional signs following from the Koszul sign rule 
$$ (f\otimes g)(x\otimes y) = (-1)^{{\tilde g}{\tilde x}} f(x) \otimes g(y)$$
where ${\tilde g}$, ${\tilde x}$ denote the degrees of $g$ and respectively $x$.
Finally, note that an associative DG algebra is an $A_\infty$-algebra in which all 
higher products $m_n$, $n\geq 3$ vanish, and $m_2$ is associative. 
At the opposite end of the spectrum, we have minimal $A_\infty$-algebras, which are 
characterized by $m_1=0$. 

In our case, one can define a minimal $A_\infty$ structure on $H$ by applying the 
construction of \refs{\SAM,\Poli,\Polii}.
Note that the restriction 
$Q|_{\hbox{Im} Q^\dagger}: \hbox{Im}(Q^\dagger) \ra \hbox{Im}(Q)$ is invertible, and let 
$Q^{-1}$ denote the inverse. Also, let $\pi : \CV \ra \hbox{Im}(Q)$ denote the projection operator 
defined by the Hodge decomposition \suppotB,\ and set $\delta = Q^{-1}\pi: \CV \ra 
\hbox{Im}(Q^\dagger)$. 
Then one first defines the multilinear maps \SAM\ 
$\lambda_n:\CV^{\otimes n} \ra \CV$, $n\geq 2$ by 
\eqn\suppotD{ \eqalign{ 
\lambda_2(a_1,a_2) & = a_1\cdot a_2 \cr
&\ \vdots \cr
\lambda_n(a_1, \ldots, a_n) & = (-1)^{n-1}\Big[\delta\lambda_{n-1}(a_1,\ldots,a_{n-1})\Big]\cdot a_n - (-1)^{n{\tilde a}_1}a_1\cdot \Big[\delta\lambda_{n-1}(a_2,\ldots, a_n)\Big]\cr
& -\sum_{k+l=n,k,l\geq 2}(-1)^{k+(l-1)({\tilde a}_1+\ldots {\tilde a}_n)} \Big[\delta\lambda_k(a_1,\ldots,a_k)\Big]\cdot
\Big[\delta\lambda_{l}(a_{k+1},\ldots, a_n)\Big],\cr}}
where ${\tilde a}$ denotes the ghost number of $a\in \CV$. 
The products $m_n :H^{\otimes n}\ra H$ are defined by 
\eqn\suppotE{
m_1 = 0,\qquad m_n =  P\lambda_n}
where $P:\CV\ra \CH\simeq H$ denotes the projection operator onto the subspace $\CH$ in the 
Hodge decomposition \suppotB.\ 

These products encode the higher order obstructions in the deformation theory 
of the holomorphic 
bundle $E$ \FukI.\ More precisely, one can formally represent the versal moduli space 
of the bundle $E$ as the zero locus of a system of formal power series of the form 
\eqn\suppotEA{ 
\sum_{n\geq 1} (-1)^{n(n+1)/2} m_n(\Phi^{\otimes n})=0}
where $\Phi\in H^{0,1}(X, \hbox{End}(E))$. 
For a rigorous construction, one has to introduce a suitable metric topology on 
$H^{0,1}(X, \hbox{End}(E))$ so that the series \suppotEA\ 
is convergent in a small neighborhood of the origin. Then the versal moduli 
space can be constructed as a germ of analytic space. Alternatively, we can work 
purely algebraically interpreting \suppotEA\ as the defining equations of a 
formal scheme \refs{\RH} (ch. II sect 9.) 
 
On physical grounds, one should be able to represent the moduli space as the 
critical locus of a holomorphic superpotential defined on the space of 
physical massless modes. 
Such a function need not a priori exist for an arbitrary $A_\infty$ structure. 
However, the $A_\infty$ structures arising in string field theory satisfy a 
cyclicity condition \GZ\ 
\eqn\suppotEB{ 
\langle \Phi_{n+1},m_n(\Phi_1, \ldots, \Phi_n)\rangle = (-1)^{n({\tilde \Phi}_1+1)} 
\langle \Phi_1,m_n(\Phi_2, \ldots, \Phi_{n+1})\rangle.}
with respect to the nondegenerate bilinear form 
\eqn\suppotEC{ 
\langle \Phi, \Psi\rangle = \int_X \Omega_X\, \hbox{Tr}\, (\Phi \Psi).} 
The superpotential is then given  by \CLiii\ 
\eqn\suppotF{ 
W(\psi_i) = \sum_{n\geq 2} {(-1)^{n(n+1)/2}\over n+1} \int_X\, \Omega_X \hbox{Tr}
\left(\Phi\ m_n(\Phi^{\otimes n})\right).}
One can check that the critical set of $W$ is determined by the equations \suppotEA.\

By construction, the products $m_n$ and the superpotential \suppotF\ 
depend on the gauge fixing data, i.e. the metric on $X$ and the hermitian structure on 
$E$. On general grounds, all such choices should be equivalent 
from a physical point of view. In technical terms, the correct statement is that two $A_\infty$ 
structures corresponding to different gauge fixing data should be quasi-isomorphic  
\refs{\GZ}. Since this is an important point, let us spell out some details here. 
An $A_\infty$-morphism between two $A_\infty$ structures 
$m_n^\prime\,:\, V^{\otimes n} \ra V$ and $m_n\,:\, V^{\otimes n} \ra V$
is specified by a sequence of maps $f_n:V^{\otimes n} \ra V$ subject to the constraints 
\eqn\suppotH{ 
\sum (-1)^{r+st}f_u\left(\II^{\otimes r} \otimes m_s'\otimes \II^{\otimes t}\right) 
= \sum (-1)^\sigma m_r \left(f_{i_1}\otimes f_{i_2} \otimes f_{i_2} \otimes 
\cdots \otimes
f_{i_r}\right).}
The sum in the left hand side of \suppotH\ runs over all decompositions $n=r+s+t$, 
and $u=r+t+1$. The sum in the right hand side runs over all $1\leq r \leq n$ and 
all decompositions 
$n=i_1+\cdots + i_r$, and $\sigma = (r-1)(i_1-1) +(r-2)(i_2-1) + \cdots + 
2(i_{r-2}-1)+(i_{r-1}-1)$. By spelling out the first condition, 
we find that $f_1m_1'=m_1 f_1$, 
that is $f_1$ is a morphism of complexes. The second condition implies that $f_1$ commutes 
with multiplication $m_2$ up to a homotopy transformation defined by $f_2$ and so on. 
A morphism $\{f_n\}$ is called a quasi-isomorphism if $f_1$ is a quasi-isomorphism of complexes
i.e. it induces an isomorphism in cohomology. Moreover a morphism $\{f_n\}$ is a 
homotopy equivalence 
if it admits an $A_\infty$-inverse. One can prove that an $A_\infty$ morphism is a 
homotopy equivalence if and only if $f_1:V \ra V$ is the identity \FukI.\ In particular, 
any quasi-isomorphism of minimal $A_\infty$ structures is a homotopy equivalence. 

The physical applications of these concepts to string field theory have been explained in \refs{\GZ,\CLiii}. 
In our particular case, we have a topological open string field theory specified 
by the holomorphic 
Chern-Simons functional \suppotA.\ The off-shell open string states form a DG-algebra $(\CV,Q)$ with respect to the boundary operator product expansion which reduces to multiplication of 
differential forms. The higher order products $m_n:H^{\otimes n}\ra 
H$ define a minimal $A_\infty$ structure on the space of massless fields, 
which is quasi-isomorphic 
to the original DG-algebra. Since the later structure is minimal, $(H, m_n)$ is called a  
minimal model of $(\CV, Q)$. 
To understand the physical content of this statement, note that 
one can construct a non-associative string field theory of massless modes with action 
functional \suppotF.\ The quasi-isomorphism of $A_\infty$-structures implies that the 
two open string field 
theories are physically equivalent \GZ.\ In particular it was shown in \CLiii\ that the 
two theories have the same local moduli space of vacua. 

Returning to the dependence of higher products on gauge fixing data, one can show 
that two different choices 
result in homotopic equivalent minimal $A_\infty$ structures $(H, m_n)$, $(H, m_n^\prime)$. 
As long as we are working in the framework of topological open string field theory, there is no 
preferred choice, and one can describe the same physics using any minimal model. 
In particular, one may have different superpotentials \suppotF\ describing the same 
moduli space. 

Although conceptually clear, the above construction is not very suitable for explicit 
computations. The main difficulty resides in the fact that the operator 
$\delta$ must be written in terms of Green's functions for the Laplace operator 
on Calabi-Yau manifolds. Moreover, we are interested in holomorphic branes 
supported on curves in $X$ rather than vector bundles $E\ra X$. In order to construct 
the string field theory in that case, we have to employ a graded version of 
Chern-Simons theory \DED\ associated to a locally free resolution of the curve. 
In principle higher products can be similarly constructed for graded Chern-Simons 
theory, but the computations would be practically intractable. 
Some explicit calculations for ${\bf A}$-branes on tori can be found in 
\refs{\LR}, but the present situation is much more complicated. Another 
computation of a higher order product using linear sigma model techniques 
has been performed in \DGJT.\ 
In the next subsection we will discuss an alternative approach to this 
problem based on the Landau-Ginzburg description of holomorphic branes. 

\subsec{Higher Products in Landau-Ginzburg Orbifold Categories}

Note that in principle we do not need the whole Hodge Theory machinery in order 
to construct a minimal model of a given DG-algebra $(\CV,Q)$ \SAM.\ The essential 
elements of this construction are a 

$i)$ a linear subspace $\CH\subset \CV$ of cohomology representatives, and 

$ii)$ an odd linear map $\delta : \CV \ra \CV$ mapping $\CH$ to itself so that 
$P=\II-[Q,\delta] : \CV\ra \CV$ is a projector onto $\CH$. 

\noindent 
Given this data, the formulae 
\suppotD,\ \suppotE\ define an $A_\infty$-structure on $H\simeq \CH$. This structure is 
quasi-isomorphic to the original DG-algebra $(\CV,Q)$ if $\delta^2=0$ and 
$P\delta =0$ \Polii.\ In the geometric phase discussed above it is hard to imagine 
an abstract non-metric dependent gauge fixing condition. 
However standard decoupling arguments \BDLR\ suggest that the topological boundary 
correlators should be independent of K\"ahler moduli. Therefore it would be very 
interesting to find an alternative construction of higher products depending only 
on complex structure moduli. 

The main point of the present paper is to exhibit such an explicit construction 
for higher products in the Landau-Ginzburg phase. 
At the Landau-Ginzburg orbifold 
point {\bf B}-twisted branes are realized as equivariant matrix factorizations of the 
LG potential.
In order to fix ideas suppose we are given such a brane  
$\op = \Big(\xymatrix{P_1 \ar@<1ex>[r]^{p_1}& P_0\ar@<1ex>[l]^{p_0}\\},R_1,R_0\Big)$, 
where $R_1, R_0$ are representations of the orbifold group $G$.
For future reference it is helpful to represent the free modules $P_1,P_0$ as 
$P_1= E_1\otimes_\IC \IC[x_0,\ldots,x_n]$, 
$P_0= E_0 \otimes_\IC \IC[x_0,\ldots, x_n]$,
where $E_1,E_0$ are complex vector spaces.
The endomorphisms of $\op$ are defined by the cohomology of the $G$-equivariant 
$\IZ/2$ graded differential algebra \complexA.\ In the following we will denote
this algebra by $(\CC,D)$. We will also use the notation $\CC^{1,0}$ for the 
degree one and respectively zero components.
In conclusion, we are presented with a differential graded algebra 
$(\CC,D)$, albeit in a $\IZ/2$ graded 
equivariant form. 

In order to obtain the gauge fixing data $(i)$ and $(ii)$, first note that $\CC$ is a free 
graded $R$-module (where $R$ is the ring of polynomials in $n+1$ variables,) 
and $D$ is an $R$-module homomorphism. Therefore one can find a set of representatives 
of cohomology classes using algebraic methods, such as Gr\"obner bases of ideals. 
This step is especially easy if the matrix factorization $\op$ is a tensor product 
of one and two variable blocks as in section two. Therefore we can grant the existence of a finite-dimensional subspace $\CH\subset \CC$ of cohomology representatives. 

To proceed next, consider the short exact sequences of graded $R$-modules 
\eqn\shortA{\eqalign{ 
& 0 \ra \CZ\ra \CC {\buildrel Q \over \ra} \CB^{+1}\ra 0\cr
& 0 \ra \CB \ra \CZ \ra H\ra 0,\cr}}
where $H$ denotes the graded cohomology space of $D$. 
We can also regard \shortA\ as exact sequences of infinite dimensional complex vector spaces. 
Note however that since we are dealing with polynomials in several variables, we will only have 
to consider linear combinations of finitely many basis elements. 
A short exact sequence of complex vector spaces is always split, therefore we can choose 
splittings $\gamma:\CB\ra \CC$ and $\rho:H\ra \CZ$ so that we have the direct sum 
decompositions 
\eqn\splittingA{ 
\CC = \CZ \oplus \hbox{Im}(\gamma), \qquad 
\CZ = \CB \oplus \hbox{Im}(\rho).}
Note that $\hbox{Im}(\rho)$ is in fact the subspace $\CH$ of cohomology representatives 
introduced in the last paragraph. Therefore we can write $\CC$ as a direct sum 
\eqn\splittingB{ 
\CC = \CB \oplus \CH \oplus \hbox{Im}\gamma.}
By construction, the projection $\pi_{\hbox{Im}(\gamma)}: \CC \ra \hbox{Im}(\gamma)$ 
is given by $\pi_{\hbox{Im}(\gamma)} = \gamma\pi_\CB D$ where $\pi_\CB:\CC \ra \CB$ 
denotes the projection onto $\CB$. Also by construction, $D\gamma=\II$. 
Let $\delta: \CC \ra \CC$ be given by the composition 
$\delta = \gamma \pi_\CB$. Then one can easily compute 
\eqn\splittingsB{ \eqalign{
\II-\left(D\delta + \delta D\right) & = \II - D\gamma \pi_\CB - \gamma \pi_\CB D \cr
& = \II - \pi_\CB - \pi_{\hbox{Im}(\gamma)} = \pi_\CH,\cr}}
where $\pi_\CH : \CC \ra \CH$ denotes the projection onto $\CH$. 
Since $\delta$ also preserves $\CH$, we can conclude that we have all the 
required ingredients for the construction of higher order products. In orbifold theories this construction needs to be performed in an equivariant setting, but this brings in no additional complications.

The discussion is of course too abstract at this point since one has to make explicit choices
of splittings in order to perform concrete computations. However, note that the D-brane 
moduli space is determined by higher products 
$m_n(\Phi^{\otimes n}) = \pi_\CH\lambda_n(\Phi^{\otimes n})$ 
evaluated on cohomology representatives $\Phi$. For tensor product matrix 
factorizations one can obtain a canonical set of cohomology representatives by taking 
tensor products of one and/or two variable morphisms. Moreover, the splittings 
$\gamma, \rho$ can be chosen to be compatible with the tensor product of morphisms. 
Therefore the computation is effectively broken into one and two 
variable pieces which are very easy to handle. We will discuss concrete examples in detail 
in section five. 

Finally, note that this approach does not make use of all elements of a string field 
theory. In particular we did not have to use a nondegenerate bilinear form on the space 
of open string morphisms at any step of the construction, although such a 
form exists and is given by a residue pairing \KLIII.\ Therefore the resulting 
$A_\infty$ coefficients are not guaranteed to satisfy the cyclicity property, and 
one cannot a priori write down a superpotential. This will be the case with the 
examples discussed in section five. However, all physical information 
is encoded in the $A_\infty$ structure up to homotopy. Therefore it suffices to find a 
homotopy transformation which makes a superpotential manifest.  
Such transformations will also play a key role in the comparison of D-brane moduli 
spaces between the Calabi-Yau and Landau-Ginzburg phase. In principle one could 
try to construct a string field theory using the residue pairing of 
\KLIII\ and define cyclic $A_\infty$-products. Such an approach would be more complicated 
at the computational level, so we will leave it for future work.

\newsec{Complex Structure Deformations} 

In this section, we consider 
both open and closed string marginal perturbations of the boundary {\bf B}-models.  
It is well known that closed string marginal perturbations 
correspond to complex structure deformations in the geometric phase. Moreover, they
are exactly marginal since complex structure deformations of Calabi-Yau threefolds 
are unobstructed. However we can have disc level couplings between bulk and boundary 
marginal operators which result in a nontrivial dependence of the open string 
superpotential on the closed string moduli. It was shown in \HLL\ that such effects 
deform the strong $A_\infty$-structure associated to a boundary CFT to a weak 
$A_\infty$ structure depending on closed string moduli. Here we will develop a constructive 
approach to these deformations by extending the above considerations to 
families of Landau-Ginzburg models. 

Let us first discuss the geometric situation. An open-closed topological {\bf B}-model
is determined in our case by a pair $(X,F)$ where $X$ is a Calabi-Yau threefold and 
$F$ is a rational curve on $X$. Therefore the local moduli space of 
open-closed TFTs is isomorphic to the versal deformation space $\CP$ of the pair 
$(X,F)$. The forgetful map $(X,F)\ra X$ induces a projection $\pi:\CP \ra \CM$ from 
$\CP$ to the versal deformation space $\CM$ of $X$. 

In the following we will consider only 
Calabi-Yau hypersurfaces $X$ in weighted projective spaces. We also restrict to 
complex structure deformations corresponding to linear deformations of 
the defining equation of $X$ in the ambient toric variety.
Therefore let us consider an $m$-parameter
family $\CX\ra T$ of Calabi-Yau hypersurfaces in weighted projective space 
$WP^{w_0,\ldots, w_n}$ parameterized by the linear space $T=\IC^m$. The total space 
$\CX$ can be regarded as a hypersurface 
in $WP^{w_0,\ldots, w_n} \times T$ defined by the equation 
\eqn\cpxmodA{ 
W_{LG}(x_0,\ldots,x_n)+t_1G_1(x_0,\ldots, x_m) +\ldots + t_mG_m(x_0,\ldots, x_m) =0}
where $G_1(x_i),\ldots,G_m(x_i)$, $i=0,\ldots, n$ are  quasihomogeneous polynomial perturbations.  
By eventually restricting to an open 
subset of $T$, we can find a classifying map $\kappa: T \ra \CM$ for the 
family $\CX/T$. Our problem is to determine the restriction $\CP_T=\CP\otimes_\CM T$ 
of the versal moduli space of pairs to $T$. Loosely speaking this is the local 
moduli space of pairs $(X,F)$ so that $X$ is a point in $T$. This point of view 
is more convenient for practical applications since we can avoid working with 
large numbers of moduli by choosing $T$ to be a low dimensional 
slice in the moduli space of $X$. 

The problem can be further simplified by noting that any point $\CP_T$ is 
represented by a curve $F_t$ on some fiber of $\CX_t$ of $\CX/T$. Therefore
any deformation of the pair $(X,F)$ gives rise to a deformation of $F$ in the 
total space of the family $\CX$. Conversely, any deformation $F'$ of $F$ in 
$\CX$ must be contained in some fiber $\CX_t$ since the base $T$ is a linear space.
Hence we obtain a one-to-one map between $\CP_T$ and the versal deformation space 
of $F$ in the total space $\CX$. By construction this map is holomorphic, therefore 
the two moduli spaces are isomorphic as germs of analytic spaces. This argument 
effectively reduces the problem to deformation of curves in the higher dimensional 
space $\CX$. Our plan is to find the defining equations of the deformation 
space by extending our previous construction of higher products to families. 

Let us first analyze infinitesimal first order deformations of 
$F$ in $\CX$. The normal bundle of $F$ in $\CX$ fits in the short exact sequence 
\eqn\cpxmodB{ 
0\ra N_{F/X} \ra N_{F/\CX} \ra \left(N_{X/\CX}\right)_F\ra 0}
where the last term $N_{X/\CX}$ is isomorphic to the trivial line bundle 
$N_{X/\CX}\simeq \CO_X$. Therefore $N_{F/\CX}$ is an extension of $\CO_F$ by 
the rank two bundle $N_{F/X}$. Such extensions are parameterized by 
$\hbox{Ext}^1(\CO_F, N_{F/X})\simeq H^1(F, N_{F/X})$. More precisely, consider the 
long exact sequence associated to \cpxmodB\ which reads in part 
\eqn\cpxmodC{  
0\ra H^0(F,N_{F/X}) \ra H^0(F,N_{F/\CX}) \ra H^0(F,\CO_F) {\buildrel \delta \over \ra} 
H^1(F, N_{F/X}) \ra \cdots}
The extension class is parameterized by $\delta(1) \in H^1(F,N_{F/X})$. If this class 
is trivial, the extension is split and we have $H^0(F,N_{F/\CX})\simeq H^0(F,N_{F/X}) \oplus H^0(F,\CO_F)$. 
In this case we obtain an extra infinitesimal deformation of $F$ in $\CX$ corresponding to 
infinitesimal displacements in the normal direction to the central fiber $X$. 
Such infinitesimal deformations will be called horizontal in the following. 
If $\delta(1)\neq 0$, the extension is nontrivial, and $F$ has no horizontal 
infinitesimal deformations in $\CX$. 

We can find a more effective characterization of the extension class by looking at infinitesimal 
deformations of $F$ in the ambient weighted projective space as in \AK.\ 
For simplicity we will consider a one 
parameter family of Calabi-Yau hypersurfaces $X_t$ of the form 
\eqn\cpxmodD{ 
W_{LG}(x_0,\ldots, x_n) + tG(x_0,\ldots, x_n) =0.}
Let 
\eqn\cpxmodE{
f_a(x_0,\ldots, x_n)=0,\qquad a=1,\ldots, A}
be the defining equations of $F$ in the weighted projective space. 
It follows from Hilbert Nullstellensatz that $F$ lies on $X$ if and only if $W_{LG}$ belongs to the 
ideal generated by the $f_a$, that is if and only if 
\eqn\cpxmodF{ 
W_{LG} = \sum_{a=1}^A f_a q_a}
for some quasi-homogeneous polynomials $q_a(x_0,\ldots, x_n)$. 
A first order deformation $F_t$ of $F$ in the ambient weighted projective space is 
given by an infinitesimal perturbation of the equations \cpxmodE\
\eqn\cpxmodG{ 
f_a(x_0,\ldots, x_n)+tg_a(x_0,\ldots, x_n)=0.}
Here $t$ is an infinitesimal first order parameter $t^2=0$. The deformed curve \cpxmodD\ 
lies on the deformed hypersurface \cpxmodD\ if and only if we can write 
\eqn\cpxmodH{ 
W_{LG} + tG= \sum_{a=1}^A(f_a+tg_a)
(q_a +t q_a')}
for some quasi-homogeneous polynomials $q_a'(x_0,\ldots, x_n)$. 
Using the condition $t^2=0$ we find  
\eqn\cpxmodI{ 
G= \sum_{a=1}^A \left(q_a'f_a+ g_aq_a\right).}
It follows that $F_t$ lies on $X_t$ if and only if $G$ belongs to the ideal generated by 
$(f_a, q_a)_{a=1,\ldots A}$. However note that $F_t$ lies on $X_t$ if and only if the extension 
\cpxmodB\ is trivial according to the above paragraph. Therefore we can conclude that the extension
\cpxmodB\ is trivial if and only if the image ${\overline G}$ of $G$ in the quotient ring 
$\IC[x_0,\ldots, x_n]/(f_a, q_a)$ is trivial. 

Let us now move on to the Landau-Ginzburg phase. According to section two, the complete 
intersection \cpxmodE\ corresponds to a matrix factorization of $W_{LG}$ of the form 
\eqn\cpxmodJ{
\of = \bigotimes_{a=1}^A
\left(\xymatrix{\IC[x_0,\ldots, x_n] \ar@<1ex>[r]^{f_a}& \IC[x_0,\ldots,x_n] 
\ar@<1ex>[l]^{q_a}\\}\right).}
Note that the cokernel of this factorization is isomorphic to the structure ring 
of $F$ as $\IC[x_0,\ldots,x_n]$-modules. 

The family $\CX$ corresponds to family of Landau-Ginzburg models with superpotential 
\cpxmodA.\ In order to construct a Landau-Ginzburg model for the curve $F$ embedded in the 
total space of the family we have to treat $t_1,\ldots,t_m$ as 
dynamical variables rather than parameters. Then a complete intersection $F$ in 
$WP^{w_0,\ldots, w_n}\times T$ contained in $\CX$ must be described by a matrix factorization 
of the form 
$${\overline \CF}=
\left(\xymatrix{\CF_1 \ar@<1ex>[r]^{\phi_1}& \CF_0\ar@<1ex>[l]^{\phi_0}\\}\right)$$  
over the polynomial ring $\IC[x_0,\ldots,x_n,t_1,\ldots,t_m]$ so that  $\hbox{Coker}(\phi_1)$ 
is isomorphic to the structure ring of $F$ as  $\IC[x_0,\ldots,x_n,t_1,\ldots,t_m]$-modules.
Such a factorization can be simply constructed as the tensor product 
\eqn\cpxmodK{ 
{\overline \CF}=\of\otimes 
\left(\xymatrix{\IC[x_i,t_j] \ar@<1ex>[r]^{t_1}& \IC[x_i,t_j]\ar@<1ex>[l]^{G_1}\\}\right)
\otimes \cdots 
\otimes 
\left(\xymatrix{\IC[x_i,t_j] \ar@<1ex>[r]^{t_m}& \IC[x_i,t_j]\ar@<1ex>[l]^{G_m}\\}\right)}
where $i=0,\ldots,n$, $j=1,\ldots,m$.
It is straightforward to check that this factorization has the right cokernel 
proceeding as in section $6$ of \ADD.\ 

For future reference, let us show that the 
space of infinitesimal deformations of ${\overline \CF}$ in the Landau-Ginzburg category 
is isomorphic to the space 
of infinitesimal deformations of $F$ in $\CX$. 
For simplicity we restrict again to one-parameter deformations in which case we have a 
single $t$-dependent factor in \cpxmodK.\ Infinitesimal deformations 
of ${\overline \CF}$ in the Landau-Ginzburg category are parameterized by the space of  
odd endomorphisms $H^1(\OCF,\OCF)$. It is straightforward to check that there is an embedding 
\eqn\cpxmodL{0\ra H^1(\of,\of){\buildrel \otimes \II\over \ra } H^1(\OCF,\OCF)} 
defined by taking tensor products by the identity
endomorphism of the $t$-dependent factor in \cpxmodK.\ 
The main questions is to determine the cokernel of the map \cpxmodL\ and 
compare the outcome to the geometric result.  

In order to answer this question, we have to write down the $\IZ/2$ graded 
morphism complex for the object ${\overline \CF}$ and 
determine its cohomology. We have included the details of this computation in appendix A. 
The answer can be most conveniently 
formulated in terms of the differential morphism complex $(\CC,D)$ associated to the 
object $\of$.
The cokernel of \cpxmodL\ is parameterized by equivalence classes 
of pairs $(\beta,\xi) \in \CC^0 \oplus \CC^1$ subject to the conditions 
\eqn\cpxmodM{ 
D\cdot \xi = G\beta,\qquad D\cdot\beta =0.} 
The equivalence relation on such pairs is defined by $(\beta',\xi') \sim (\beta,\xi)$ if 
\eqn\cpxmodN{ 
\beta'-\beta \in \CB^0, \qquad \xi'-\xi \in \CZ^0.} 
This means that $\beta$ is a representative of a bosonic endomorphism in $H^0(\of,\of)$ and $\xi$ is a trivialization of the cocycle $G\beta$ defined up to addition of a closed element. 

For comparison to the geometric result, we have to keep in mind that
morphism spaces in the Landau-Ginzburg orbifold category correspond to 
graded sums of morphism spaces in the derived category as explained in section two, 
equation \morphA.\ Therefore $H^1(\OCF,\OCF)$ should be compared to the direct sum 
\eqn\cpxmodO{\eqalign{
\bigoplus_{k=1,3}\hbox{Ext}^k(\CO_{F/\CX},\CO_{F/\CX}) \simeq &
 H^0(F,\Lambda^1(N_{F/\CX}))\oplus H^0(F,\Lambda^3(N_{F/\CX}))\oplus \cr
& H^1(F,\Lambda^0(N_{F/\CX}))\oplus H^1(F,\Lambda^2(N_{F/\CX})).\cr}}
Let us first assume that the extension \cpxmodB\ is split, that is the image ${\overline G}$ 
of $G$ in the quotient ring $\IC[x_i]/I_F$ is trivial. Then $N_{F/\CX}\simeq 
N_{F/X}\oplus \CO_F$ and we can evaluate \cpxmodO\ 
obtaining 
\eqn\cpxmodP{\eqalign{
\bigoplus_{k=1,3}\hbox{Ext}^k(\CO_{F/\CX},\CO_{F/\CX}) \simeq &  
H^0(F, N_{F/X})\oplus H^1(F,\Lambda^2(N_{F/X})) \cr
& H^0(F,\CO_F)\oplus H^1(F,N_{F/X}).\cr}}
Note that in the right hand side of \cpxmodP\ we have 
\eqn\cpxmodQ{\eqalign{ 
& H^0(F, N_{F/X})\oplus H^1(F,\Lambda^2(N_{F/X})) \simeq \bigoplus_{k=1,3} \hbox{Ext}^k
(\CO_{F/X}, \CO_{F/X})\simeq H^1(\of,\of)\cr
&  H^0(F,\CO_F)\oplus H^1(F,N_{F/X})\simeq \bigoplus_{k=0,2} \hbox{Ext}^k
(\CO_{F/X}, \CO_{F/X})\simeq H^0(\of,\of).\cr}}
Therefore, if ${\overline G}=0$, we obtain one extra odd endomorphism for each even 
endomorphism of ${\underline \CO_F}$ in $D^b(X)$. 
Moreover, if we take $\beta=\II$ in equation \cpxmodM,\ the resulting condition 
$G=D\cdot\xi$ is equivalent to ${\overline G}=0$. This follows from the explicit form 
of the differential $D$ on the morphism complex of the factorization \cpxmodJ.\
According to equations \cpxmodM,\ \cpxmodN,\ if this condition is satisfied 
we obtain one extra fermionic endomorphism $\Xi$ 
of ${\overline \CF}$ for each bosonic endomorphism $\beta$ of $\of$. 
In conclusion if $\og=0$, there is a precise one-to-one correspondence between 
endomorphisms of $\OCF$ in the Landau-Ginzburg category and derived endomorphisms of 
$\CF$.
If ${\overline G}\neq 0$, we can show by a similar reasoning that the two results also agree.
This is a remarkable confirmation of our construction. 

In the remaining part of this section we will determine the versal deformation space 
of $\OCF$ by constructing higher products in the Landau-Ginzburg orbifold category of the 
deformed superpotential \cpxmodD.\ Then we will show that this construction is encoded 
in a simple deformation of the $A_\infty$ structure associated to the initial object 
$\of$. 

From now on we will work under the assumption ${\overline G}=0$, so that we have exactly 
one horizontal deformation corresponding to the cohomology class $\Xi$ found in 
appendix A. In principle higher order products on the endomorphism algebra 
$H(\OCF,\OCF)$ can be constructed following the steps described in section 3.2, 
since we are now reduced to a similar problem in a higher dimensional set-up. 
Namely we have to fix a set ${\cal K}$ of cohomology representatives in the morphism complex 
$(\IH(\OCF,\OCF),\CD)$ and find an odd operator $\Delta : \IH(\OCF,\OCF)\ra \IH(\OCF,\OCF)$ 
satisfying conditions $(i)$, $(ii)$ of section 3.2. Then the higher products are determined 
by the equations \suppotD,\ \suppotE.\ 

The infinitesimal deformations of $\CF$ we are interested in are precisely those 
corresponding to infinitesimal deformations of $F$ in $\CX$. They can be parameterized by 
$\alpha \Xi +\Lambda$  where $\Lambda$ is a cocycle of the form 
$\Phi\otimes \II$ with $\Phi$ a cohomology representative for the morphism 
complex $(\IH^1(\of,\of), D)$ as in \cohomrep, and $\alpha$ is a complex parameter.
Recall that the $\psi_i$ in \cohomrep\ parametrize infinitesimal deformations of the brane $\of$ which correspond to deformations of the curve $F$ in the fixed threefold $X$. 
The complex parameter $\alpha$ parameterizes infinitesimal horizontal deformations 
along the base of the family. Therefore it should be regarded as a closed string modulus. 
In this construction closed string moduli are realized as open string moduli in a higher 
dimensional model. 

As shown in appendix A, all products of the form $\Xi\Lambda$, 
$\Lambda\Xi$, $\Xi^2$ and $\Lambda^2$ lie in the image of the tensor product map 
\eqn\cpxmodS{ 
0\ra \IH^0(\of,\of){\buildrel \otimes \II\over \ra} \IH^0(\OCF,\OCF).}
Then the problem can be considerably simplified by choosing 
$\Delta$ so that it restricts to $\delta \otimes \II$ on $\IH^0(\of,\of)\otimes \II
\subset \IH^0(\OCF,\OCF)$. (Recall that $\delta : \IH^0(\of,\of)\ra \IH^1(\of,\of)$ 
is the odd morphism used in the construction of the products $m_n$ in the previous 
section.)
Making such a choice for $\Delta$, it follows from the defining relations \suppotD,\ \suppotE\ that the products $\mu_n : H^1(\OCF,\OCF)^{\otimes n} \ra H^0(\OCF,\OCF)$ take values in 
$H^0(\of,\of)\otimes \II$ when evaluated on elements of the form $\alpha \Xi+\Lambda$. 

The equations of the moduli space can be formally 
written as 
\eqn\cpxmodT{ 
\sum_{n\geq 1} (-1)^{n(n+1)/2} \mu_n\left((\alpha \Xi+\Lambda)^{\otimes n}\right) =0.}
By expanding in powers of $\alpha$, and collecting the terms, this equation can be rewritten 
in the form 
\eqn\cpxmodU{ 
\sum_{n\geq 0} (-1)^{n(n+1)/2} m_n^\alpha (\Lambda^{\otimes n})=0}
where 
\eqn\cpxmodV{ 
m_n^\alpha(\Lambda^{\otimes n}) = \sum_{k_0,\ldots,k_n} (\pm)\alpha^k
\mu_{n+k}((\Xi)^{\otimes k_0}, \Lambda, (\Xi)^{\otimes k_1}, \Lambda,\ldots, 
\Lambda, (\Xi)^{\otimes k_n})}
This is a familiar construction in the theory of $A_\infty$ algebras \FukII\ Prop. 13.40 
(see also \refs{\HLL,\CLiv} for some applications to physics.) 
Given any strong $A_\infty$ algebra $m_n:V^{\otimes n} \ra V$, and a cochain $b\in V$ one
can define a series of deformed products 
\eqn\cpxmodW{ 
m_n^b(a_1,\ldots,a_n) = \sum_{k_0,\ldots,k_n}(\pm) 
\mu_{n+k}(b^{\otimes k_0}, a_1, b^{\otimes k_1}, a_2,\ldots, 
a_n, b^{\otimes k_n})}
which form a structure of weak $A_\infty$ algebra. In particular $m_0^b$ may be nonzero. 
In this context this construction encodes the behavior of open string higher products 
under closed string complex structure deformations. 
Note that all products in the right hand side of \cpxmodV\ take values in 
$H^0(\of,\of)\otimes \II$, 
therefore they can be regarded as linear maps 
$m_n^\alpha :H^1(\of,\of)^{\otimes n} \ra H^0(\of,\of)$.
This is a deformation of the original $A_\infty$ structure. Although we have focused on 
odd cohomology classes, one can write similar formulae for even cohomology classes paying 
special attention 
to signs. We will not give more details here. 
Note also that multivariable closed string deformation can be treated along the same lines. 

\newsec{A Concrete Example -- Lines on The Fermat Quintic} 

In this section we apply our construction to lines on the Fermat quintic threefold $X$. 
The Landau-Ginzburg orbifold is given by the superpotential $W_{LG}= x_0^5 +\ldots +x_4^5$
with a $G=\IZ/5$ orbifold projection $x_i\ra \omega x_i$, $\omega^5=1$. We consider lines $F$ 
on $X$ determined by the linear equations 
\eqn\quintA{ 
x_0-\eta_{01}x_1=0,\qquad x_2-\eta_{23}x_3 =0,\qquad x_4=0}
where $\eta_{01}^5=\eta_{23}^5=-1$. 
The associated Landau-Ginzburg brane is a tensor product factorization of $W_{LG}$ of the form 
\eqn\quintB{
\of=\op^{(0,1)}_{\eta_{01}} \otimes \op^{(2,3)}_{\eta_{23}}\otimes \om^{(4)}_{1}}
in the notation of section $2\,$. 

For computational purposes, it is very convenient to write the tensor product \tensorA,\ \tensorB\ 
in terms of free boundary fermions, as explained in section $3$ of \ADD. 
We introduce a set of anticommuting variables satisfying the algebra
\eqn\anticom{
\{\pi_\alpha,\pi_\beta\}=\{\bar{\pi}_\alpha,\bar{\pi}_\beta\}=0 \quad\quad 
\hbox{and} \quad\quad \{\pi_\alpha,\bar{\pi}_\beta\}=\delta_{\alpha\beta}
\qquad \alpha, \beta=1,2,3\,.}
Then the map $f=f_1\oplus f_0: F_1\oplus F_0 \ra F_1\oplus F_0$ can be expressed as a linear combination 
\eqn\ptot{
f= (x_0 - \eta_{01} x_1) \pi_1 + Q_{\eta_{01}} \bar{\pi}_1 + 
(x_2 - \eta_{23} x_3)  \pi_2 + Q_{\eta_{23}} \bar{\pi}_2 + x_4 \pi_3 +x_4^4 
\bar{\pi}_3\,,}
where $Q_{\eta}$ has been defined below \blocksB. 
By choosing a suitable representation of the complex Clifford algebra one can recover the block matrix 
expressions for $(f_0,f_1)$.  
In this formulation the cochains in the morphism complex $\IH(\of,\of)$ can be written as linear 
combinations of monomials $\pi^I{\bar\pi}^J=\prod_{a=1}^3\pi_a^{I(a} {\bar\pi}_a^{J(a)}$ 
with polynomial coefficients  
\eqn\quintC{
\Phi= \sum_{I,J}\Phi_{I,J}(x_i)\pi^I{\bar\pi}^J,}
where $I(a), J(a)$ take values $0,1$.
The $\IZ/2$ grading is given by $\sum_{a=1}^3(I(a)-J(a))$ mod 2 and the differential 
$D$ is given by the graded commutator 
\eqn\quintD{ 
D=[f,\ ].}

In the absence of an orbifold projection, the endomorphism algebra of the morphism spaces 
$H^k(\of,\of)$ can be determined in terms of the morphisms spaces of the individual 
factors using the algebraic K\"unneth formula \ADD.\ In terms of free fermions, this means that the 
cohomology representatives of the morphism complex $\left(\IH(\of,\of),D\right)$ can be written as 
\eqn\tensorPhi{
\Phi=\prod_{\alpha=1}^3 \Phi_{\alpha} = \prod_{\alpha=1}^3 \sum_{I(\alpha),J(\alpha)}\Phi_{I(\alpha),J(\alpha)} 
\pi_{\alpha}^{I(\alpha)}{\bar\pi_{\alpha}}^{J(\alpha)}.}
The $\Phi_{\alpha}$ are cohomology representatives for physical morphisms of the individual factors. For future reference we denote by $D^{(\alpha)}$ the corresponding differentials. In the presence of an orbifold projection, we have to project onto invariant morphisms. 
An efficient implementation of the orbifold projection can be achieved by 
assigning morphisms charges corresponding to irreducible representations of the orbifold group $G=\IZ/5$.  
Then we keep only morphisms of charge $0$ mod 5.    
We have defined the action of the orbifold group on the fields $x_i$ such that each of them is assigned charge 
$1$. If we further require that $f$ given in \ptot\ be neutral, the charges of 
$\pi_{\alpha}$ and $\bar{\pi}_{\alpha}$ are fixed to be respectively $-1$ and $1$. 

Let us now explicitly construct the endomorphisms. We begin by studying the cohomology of $D^{(1)}$. Using the anticommutation relations \anticom,\ and making a linear change of variables 
$$y_0 = {1\over 2}(\eta_{01}^{-1} x_0 + x_1),\qquad 
y_1 = {1\over 2}( x_0 - \eta_{01} x_1),$$ 
we find that $D^{(1)}$ acts on generic bosonic and fermionic morphisms as
\eqn\Done{\eqalign{
D^{(1)}\cdot (a_1 \pi_1 \bar{\pi}_1 + b_1 \bar{\pi}_1 \pi_1 )& = (b_1-a_1) 
\left[2\,y_1 \pi_1 - \Big(y_1^4 + 10\, \eta_{01}^2\,  y_1^2 y_0^2 + 5\, \eta_{01}^4\, 
y_0^4 \Big)\bar{\pi}_1\right] \cr
D^{(1)}\cdot (s_1 \pi_1  + t_1 \bar{\pi}_1 )& = 2\,t_1\, y_1 + s_1\Big(y_1^4 + 10\, \eta_{01}^2\, y_1^2 y_0^2 + 5 \,\eta^4_{01}\, y_0^4 \Big)\,,}}
where $a_1$, $b_1$, $t_1$ and $s_1$ are polynomials in  $y_0$ and 
$y_1$. Here, our notation is as follows: the coefficients that appear in morphisms in the $k$-th factor are denoted $(a_k,\,b_k)$ for the bosons and $(s_k,\,t_k)$ for the fermions. All fermionic morphisms that satisfy the closedness condition are necessarily exact, so there are no nontrivial fermionic morphisms in the cohomology of $D^{(1)}$. There are instead nontrivial bosonic morphisms, of the form 
$$
a_1 (\pi_1 \bar{\pi}_1 + \bar{\pi}_1 \pi_1)= a_1 \II \qquad \hbox{with} 
\qquad a_1\in {\IC[y_0]/ y_0^4}\,.
$$
This result also holds for the cohomology of $D^{(2)}\,$, with $y_0$ and 
$y_1$ replaced by 
$$y_2 = {1\over 2}(\eta_{23}^{-1} x_2 + x_3),\qquad y_3 = {1\over 2}(x_2 - \eta_{23} x_3).$$
In the following for simplicity we will consider an object with $\eta=\eta_{01}=\eta_{23}\,$. 
 
Next, we consider the cohomology of $D^{(3)}$. The action on bosonic and fermionic 
morphisms is
\eqn\Dthree{\eqalign{
D^{(3)}\cdot (a_3 \pi_3 \bar{\pi}_3 + b_3 \bar{\pi}_3 \pi_3 )& = (b_3-a_3) 
\left[x_4\, \pi_3 - x_4^4 \,\bar{\pi}_3\right]\cr
D^{(3)}\cdot (s_3 \pi_3 + t_3 \bar{\pi}_3 )& = t_3\, x_4 + s_3\, x_4^4\,,\cr}}
where now all the coefficients are polynomials in $x_4\,$.
In this case we find only one bosonic morphism, the identity, and one 
fermionic morphism, of the form 
$$
\pi_3 - x_4^3\bar{\pi}_3 \,.
$$ 
Now we can construct the endomorphisms of $\of$ by taking tensor products as explained above. 
If we further project onto operators of charge $0$ mod 5, 
we are finally left with three bosonic and three fermionic endomorphisms: 
\eqn\morph{\eqalign{
\II\qquad\qquad\qquad\qquad\qquad  S &= y_0\, (\pi_3 - x_4^3\bar{\pi}_3)\cr
U = y_0^3 y_2^2 \,\II\qquad\qquad\qquad\qquad\qquad  T & = y_2\, (\pi_3 - x_4^3\bar{\pi}_3)\cr
V = y_0^2 y_2^3 \,\II \qquad\quad\qquad U\cdot S = V\cdot T &= y_0^3y_2^3 \, (\pi_3 - x_4^3\bar{\pi}_3)\cr
}}
This agrees with the geometric result 
\eqn\quintD{\eqalign{
& \hbox{Ext}^0(\of,\of) = \IC,\qquad\hbox{Ext}^2(\of,\of) =  \IC^2\cr
& \hbox{Ext}^1(\of,\of)= \IC^2,\qquad \hbox{Ext}^3(\of,\of)=\IC,\cr}}
taking into account equations \morphA.\ 
A more refined comparison with the geometric morphisms can be obtained by constructing an explicit 
isomorphism between the derived modules \RHomA\ and \RHomB\ as in section five of \ADD.\ 
Without giving the full details here, the outcome is that the first two fermions in \morph\ 
can be identified with generators of $\hbox{Ext}^1(\of,\of)$, while the third fermion can be 
identified with a generator of $\hbox{Ext}^3(\of,\of)$.

In order to be able to compute the products $\lambda_k$, we still need to define an 
odd operator $\delta: \IH(\of,\of)\ra \IH(\of,\of)$ as explained in section three. 
Roughly speaking this involves two steps: a projection onto exact cochains followed by 
a choice of trivialization of exact cochains with respect to $D$. 
In fact, we will only need to evaluate $\delta$ explicitly on D-closed bosonic 
morphisms, which have a simple form: they are polynomials multiplying the identity. 
If a morphism of this form contains any nonzero powers of $x_4$, it is exact and 
\Dthree\ shows that we can choose a trivialization 
\eqn\deltaxfour{
\delta\big(P(y_0, \ldots y_3, x_4)\, x_4\, \II\big) = P(y_0, \ldots y_3,x_4)\, \bar{\pi}_3 }
where $P$ denotes a generic polynomial.  If the  morphism is independent of $x_4\,$, 
but it contains any powers of the variables $y_1$ and $y_3\,$, then it is still exact 
and it can be trivialized (using \Done) as
\eqn\deltayone{
\delta\big(P(y_0, \ldots y_3)\, y_1\,\II\big) = P(y_0, \ldots y_3)\, {\bar{\pi}_1\over 2}\,,}and analogously with $y_1$ replaced by $y_3$ and $ \bar{\pi}_1$ replaced by $\bar{\pi}_2\,$.
Finally, a polynomial that only depends on the variables $y_0$ and $y_1\,$  is exact if it is of degree $4$ 
or higher in either $y_0$ or $y_2\,$.
In this case
\eqn\deltayzero{
\delta\big( P(y_0,y_2)\, y_0^4\, \II\big) = P(y_0, y_2) \left(-{\eta\over 5}\pi_1 + 
\left(\eta^3\, y_1 y_0^2 + {\eta\over 10} y_1^3\right) \bar{\pi}_1\right)}
and similarly for $x_2^4\,$.
In principle we can extend the construction of $\delta$ to more general cochains by linearity, 
but we do not need to do this here. 
 
Consider an arbitrary linear combination of fermionic endomorphisms of the form 
\eqn\quintE{
\Phi = \psi_1\,S+\psi_2\,T = (y_0\,\psi_1 + y_2\,\psi_2) (\pi_3 - x_4^3\bar{\pi}_3).}
We next determine the products $\lambda_n,m_n$ introduced in section $3$ using the defining relations 
\suppotD,\ \suppotE\ and the anticommutation relations \anticom.\ 
At the first step we find 
\eqn\quintF{
\lambda_2 (\Phi^{\otimes 2}) = \Phi^2 = -(y_0\,\psi_1 + y_2\,\psi_2)^2\, x_4^3.}
We see from \deltaxfour\ that this is exact. 
Thus $\delta \lambda_2(\Phi^{\otimes 2}) = -(y_0\,\psi_1 + y_2\,\psi_2)^2\, x_4^2\,\bar{\pi}_3$ and $m_2=0$. 
Similar calculations yield the following formulae for the remaining higher products up to order fifteen. 
For example we have 
\eqn\quintG{\eqalign{
\lambda_3(\Phi^{\otimes 3}) &= \{a,\delta \lambda_2\}  = -(y_0\,\psi_1 + y_2\,\psi_2)^3\, x_4^2 \cr
 & m_3(\Phi^{\otimes 3}) = 0, \quad \delta \lambda_3 = -(y_0\,\psi_1 + y_2\,\psi_2)^3\, x_4 \bar{\pi}_3 \cr
\lambda_4(\Phi^{\otimes 4}) &= -\{a,\delta \lambda_3\} - \delta \lambda_2 \delta \lambda_2 = (y_0\,\psi_1 + y_2\,\psi_2)^4\, x_4 \cr
& m_4(\Phi^{\otimes 4}) = 0,\quad \delta \lambda_4 = (y_0\,\psi_1 + y_2\,\psi_2)^4\, \bar{\pi}_3 \cr
\lambda_5(\Phi^{\otimes 5}) & = \{a,\delta \lambda_4\} - \{\delta \lambda_3,\delta \lambda_2\}= (y_0\,\psi_1 + y_2\,\psi_2)^5 \cr
&m_5(\Phi^{\otimes 5})= 10\, \psi_1^3\, \psi_2^2\, y_0^3\, y_2^2 +10\, \psi_1^2\, \psi_2^3\, y_0^2\, y_2^3\, \cr
&\qquad\qquad =10\,(\psi_1^3\,\psi_2^3\,U+\psi_2\,\psi_2^3\,V).
}}
Using \deltayzero\ we can write
\eqn\quintH{\eqalign{
\delta\lambda_5(\Phi^{\otimes 5}) &= (\psi_1^5\,y_0+5 \psi_1^4\psi_2\,y_2)\left[-{\eta\over 5}\pi_1 + \left(\eta^3\, y_1 y_0^2 + {\eta\over 10} y_1^3\right) \bar{\pi}_1\right]\cr
& +(\psi_2^5\,y_2+5 \psi_2^4\psi_1\,y_0)\left[-{\eta\over 5}\pi_2 + \left(\eta^3\, y_3 y_2^2 + {\eta\over 10} y_3^3\right)\bar{\pi}_2\right]
}}
Since $\delta \lambda_5(\Phi^{\otimes 5})$ anticommutes with $\Phi$, the next non-zero product is 
\eqn\quintI{\eqalign{
\lambda_{10}(\Phi^{\otimes 10})=\delta\lambda_5\cdot\delta\lambda_5 = &-{\eta\over 5}
(\psi_1^5\,y_0+5 \psi_1^4\psi_2\,y_2)^2\left(\eta^3\, y_1 y_0^2 + {\eta\over 10} y_1^3\right) \cr
& -{\eta\over 5}(\psi_2^5\,y_2+5 \psi_2^4\psi_1\,y_0)^2\left(\eta^3\, y_3 y_2^2 + {\eta\over 10} y_3^3\right).
}}
This projects to zero in cohomology because $y_1$ and $y_3$ are exact. The next non-zero product is 
$\lambda_{15}(\Phi^{\otimes 15})$ and this time we obtain a non trivial element in cohomology:  
\eqn\quintJ{
m_{15}(\Phi^{\otimes 15})=\Big({3\over 10}\,\psi_1^{13}\,\psi_2^2+{5\over 2}\,\psi_2^{12}\,\psi_1^3\Big)\,U+\Big({3\over 10}\,\psi_2^{13}\,\psi_1^2+{5\over 2}\,\psi_1^{12}\,\psi_2^3\Big)\,V}
Proceeding similarly, one could compute in principle products of arbitrarily high order. 
Next we perform similar computations in the presence of closed string perturbations using the 
method developed in section four. 

\subsec{Lines on a Perturbed Fermat Quintic} 

We now add a one-parameter perturbation to the superpotential of the form 
\eqn\quintJA{
W_{LG}(x_0,\ldots, x_4,t) = x_0^5+\ldots+x_4^5 + t\, G(x_0,\ldots, x_4)\,,}
with $G(x_i)$ a homogeneous polynomial of degree $5$. As explained in section four, the complex parameter $t$ 
should be promoted to a dynamical variable, obtaining a higher dimensional Landau-Ginzburg model. 
The object $\OCF$ introduced in section four takes the form 
\eqn\quintK{
\op^{(0,1)}_{\eta} \otimes \op^{(2,3)}_{\eta}\otimes \om^{(4)}_{1}\otimes 
\left(\xymatrix{\IC[x_i,t] \ar@<1ex>[r]^{t}& \IC[x_i,t]\ar@<1ex>[l]^{G(x_i)}\\}\right)\,. 
}
Although this object is constructed as a tensor product, in this case it is not possible to determine its endomorphisms applying the K\"unneth formula. 
The reason is that, taken separately, the perturbation $tG(x_i)$ does not have 
isolated critical points. Therefore the space of endomorphisms of the last factor in 
\quintK\ is infinite dimensional. Instead one has to perform a direct analysis as in appendix A. 
In addition to fermionic morphisms obtained by multiplying $\IH^1(\of,\of)$ by the identity, we obtain one extra generator $\Xi$ if the perturbation $G$ is a trivial bosonic cochain in the complex $(\IH(\of,\of),D)$. Moreover, $\Xi$ corresponds to a horizontal deformation of 
the curve $F$ in the total space of the family \quintJA.\ There are other generators as well, but they correspond to higher $\hbox{Ext}$ elements in the geometric phase. 

In terms of free fermions, the differential $D$ in the deformed theory is defined as in \quintD, with $f$ replaced by
$$
f_t = f + t\,\pi_4 + G(x_i)\,\bar{\pi}_4.
$$
Here we have introduced a fourth pair of fermionic variables $(\pi_4,\,\bar{\pi}_4)$ so that now the indices $\alpha,\,\beta\,$ in \anticom\ run from $1$ to $4$. These have orbifold charge zero. In this notation, 
$$
\Xi = \xi-\pi_4,
$$ 
with $D\cdot\xi=G$. We should note that this additional morphism  has by construction orbifold charge zero and it is thus a true endomorphism in the orbifolded theory. Note that $\Xi$ does not depend explicitly on $t$, although $t$ is a dynamical variable. However, one can check that an infinitesimal deformation of $\OCF$ parameterized by $\alpha\Xi$ is equivalent in the geometric phase to shift $t\ra t+\alpha$. Therefore $\alpha$ should be thought of as a closed string modulus, as also explained in section $4\,$. 

Let us now consider an example and compute the products $m_n$ for a particular choice of perturbation. We take
$$
G=x_0\, x_2^2\,x_4^2\,,
$$ 
and in this case, one can check using \deltayzero\ that the new fermionic endomorphism is of the form
\eqn\quintN{
\Xi= x_0\,x_2^2\,x_4\,\bar{\pi}_3-\pi_4.}
The deformed products $m_n^\alpha \left(\Phi^{\otimes n}\right)$ can be determined recursively as explained in section $4$. The contributions to the $m_n^\alpha$ that are indepedent of $\alpha$ remain the same as before. The nonzero $\alpha$-dependent contributions up to order nine, are 
\eqn\quintO{\eqalign{
m_2^\alpha(\Phi^{\otimes 2})&= 
\alpha\,\eta^3\Big(\psi_1^2\,U+2\psi_1\psi_2\,V\Big)\cr
m_4^\alpha(\Phi^{\otimes 4}) &=
-\alpha^2\,{\eta\over 5}\Big(2\,\psi_2^3\psi_1\,U+\psi_2^4\,V\Big)\cr 
m_6^\alpha(\Phi^{\otimes 6}) &= \alpha^3\,{\eta^4\over 100}
\Big(\psi_2^6\,U+2\,\psi_2^2\psi_1^4\,V\Big).\cr}}
In principle, such computations can be performed up to arbitrarily high order. 

At this point we should think about the physical meaning of our results. 
As anticipated in the concluding 
remarks to section $3\,$, the moduli space equations 
$$ 
\sum_{n\geq 1} (-1)^{n(n+1)/2} m_n(\Phi^{\otimes n})=0
$$ 
are not integrable i.e. the resulting analytic space cannot be written as the critical 
locus of a superpotential $W$. This problem occurs because with our choice of $\delta$, 
the resulting higher products do not satisfy the cyclicity condition with respect to a 
nondegenerate bilinear form on open string states. However the $A_\infty$ products do encode 
the physical information needed for writing down a D-brane superpotential, except that 
we have to perform homotopy transformations to make it manifest. We will do this explicitly 
after constructing D-brane moduli spaces in the geometric phase. Then we will show that the 
geometric and non-geometric moduli spaces agree up to homotopy transformations. 
 
\subsec{The Hibert Scheme of Lines on The Quintic}

{\it a) Undeformed Case}

We will first consider the Hilbert scheme of lines on the Fermat quintic $X$. 
According to \AK\ the Hilbert scheme of lines on $X$ has  
fifty irreducible components, each component being isomorphic to a quintic curve in $\IP^2$. 
The corresponding families of lines on $X$ span fifty cones $C_{ij}$ 
determined by the equations 
\eqn\hilbB{ 
x_i-\eta_{ij} x_j=0, \qquad \sum_{k=0,k\neq i,j}^4x_k^5 =0}
where $i\neq j$ and $\eta_{ij}^5=-1$. There are ten possible choices for the pair $(i,j)$ and five 
independent choices for $\eta_{ij}$ giving fifty components as stated above. 
Note that each of these components is a cone with apex $V_{ij}=\{x_i=\eta, x_j=1,x_k=0, k\neq i,j\}$
over the quintic curve $\sum_{k=0,k\neq i,j}^4x_k^5 =0$ in the projective plane $x_i=x_j=0$. 
These cones are not disjoint; for example $C_{ij}$ and $C_{kl}$ with $(i,j)\neq (k,l)$  
(as unordered pairs) share a common line
\eqn\hilbC{
x_i-\eta_{ij} x_j =0, \qquad x_k-\eta_{kl} x_l=0, \qquad x_m=0\ \hbox{for}\ m\neq i,j,k,l.}
Without loss of generality, we can take $i=0, j=1, k=2, l=3, m=4$ in \hilbC.\ We denote 
by $F$ the line 
\eqn\hilbD{ 
x_0-\eta_{01}x_1=0,\qquad x_2-\eta_{23}x_3 =0, \qquad x_4=0.}
Our goal is to determine the local analytic type of the Hilbert scheme at the point $F$. 

The Hilbert scheme of lines on a quintic hypersurface can be represented as a complete 
intersection in the Grassmannian $G(2,5)$ of lines in $\IP^4$ \JH.\ Here we would like to write down local equations on the Hilbert scheme in a suitable affine open subset $U\subset G(2,5)$ containing $F$. There is a standard construction for such affine open subsets \refs{\GH,\JH}. 
Lines in $\IP^4$ are in one-to-one correspondence with two-planes in $\IC^5$. In particular 
the line \hilbC\ is the projectivization of the two plane $\Lambda_F$ spanned by the 
vectors 
\eqn\hilbD{
v_1 = \left[\eta_{01},1,0,0,0\right],\qquad 
v_2 = \left[0,0,\eta_{23},1,0\right].}
Choose a complementary $3$-plane $\Lambda_F^\circ$ which intersects $\Lambda_F$ only at the origin. 
The open subset $U$ is defined to be the subset of $G(2,5)$ consisting of all planes $\Lambda$ 
complementary to $\Lambda_F^\circ$. This means that $\Lambda \cap \Lambda_F^\circ =\{0\}$, 
and $\Lambda\oplus \Lambda_F^\circ=\IC^5$. Moreover, one can choose basis vectors $v_1(\Lambda), 
v_2(\Lambda)$ for each plane $\Lambda\in U$ of the form 
\eqn\hilbE{ 
v_{1,2}(\Lambda) = \Lambda \cap (\Lambda_F^\circ + v_{1,2})}
This defines a system of affine coordinates on $U$. 
We take $\Lambda^\circ_F$ to be generated by 
\eqn\hilbF{ 
v_3 = [0,1,0,0,0],\qquad v_4=[0,0,0,1,0], \qquad v_5 = [0,0,0,0,1].} 
Then it follows by some elementary linear algebra that $U$ is isomorphic to $\IC^6$, 
and one can define affine coordinates $(\alpha_1,\beta_1,\gamma_1,\alpha_2,\beta_2,\gamma_2)$ 
on $U$ by choosing 
\eqn\hilbFF{ 
v_1(\Lambda) = [\eta_{01},\alpha_1, 0, \beta_1, \gamma_1],\qquad 
v_2(\Lambda) = [0,\alpha_2,\eta_{23},\beta_2,\gamma_2].}
Therefore an arbitrary plane $\Lambda\in U$ has the following parametric form
\eqn\hilbG{ 
[x_0(u,v), \ldots, x_4(u,v)] = [\eta_{01}u,\alpha_1 u +\alpha_2 v, \eta_{23}v, 
\beta_1 u + \beta_2 v , 
\gamma_1 u + \gamma_2 v ]}
where $(u,v)\in \IC^2$ are complex parameters. 
The condition for a line parameterized by a point 
$(\alpha_1,\beta_1,\gamma_1,\alpha_2,\beta_2,\gamma_2)$
to lie on the Fermat quintic is that 
\eqn\hilbH{ 
-u^5+(\alpha_1 u + \alpha_2 v)^5 - v^5 +(\beta_1 u + \beta_2 v)^5 + (\gamma_1 u 
+ \gamma_2 v)^5 =0}
for any $(u,v)\in \IC^2$. 
By expanding the binomials in \hilbH,\ we find 
that the local equations of the Hilbert scheme in $U$ are 
\eqn\hilbI{ \eqalign{ 
& \alpha_1^5+\beta_1^5+\gamma_1^5-1=0\cr
& \alpha_1^4\alpha_2 + \beta_1^4\beta_2 + \gamma_1^4 \gamma_2 = 0\cr
&  \alpha_1^3\alpha_2^2 + \beta_1^3\beta_2^2 + \gamma_1^3 \gamma_2^2 = 0\cr
& \alpha_1^2\alpha_2^3 + \beta_1^2\beta_2^3 + \gamma_1^2 \gamma_2^3 = 0\cr
& \alpha_1\alpha_2^4 + \beta_1\beta_2^4 + \gamma_1\gamma_2^4= 0\cr
& \alpha_2^5+\beta_2^5+\gamma_2^5-1=0.\cr}}
We want to find the analytic type of this variety at the point $F$ given by 
$\alpha_1= 1,\beta_1=0,\gamma_1=0, \alpha_2=0, \beta_2=1, \gamma_2=0$. 
Let us first perform the coordinate change 
\eqn\hilbJ{ \eqalign{ 
& \talpha_1= \alpha_1-1, \qquad \tbeta_1=\beta_1,\qquad \tgamma_1 = \gamma_1 \cr
& \talpha_2= \alpha_2,\qquad \tbeta_2 = \beta_2 -1, \qquad \tgamma_2=\gamma_2\cr}}
so that the new coordinate system is centered at $F$. The equations \hilbI\ become 
\eqn\hilbK{\eqalign{ 
& \talpha_1 + 2 \talpha_1^2 + 2 \talpha_1^3 + \talpha_1^4 + {1\over 5} 
\left(\talpha_1^5+\tbeta_1^5+\tgamma_1^5\right)=0\cr
& \talpha_2 + \left(4\talpha_1+6\talpha_1^2+4\talpha_1^3+\talpha_1^4\right)\talpha_2 + 
\tbeta_1^4(1+\tbeta_2) + \tgamma_1^4\tgamma_2=0\cr
& (1+\talpha_1)^3\talpha_2^2 + \tbeta_1^3(1+\tbeta_2)^2 +  \tgamma_1^3\tgamma_2^2=0\cr
& (1+\talpha_1)^2\talpha_2^3 + \tbeta_1^2(1+\tbeta_2)^3 +  \tgamma_1^2\tgamma_2^3=0\cr
& \tbeta_1 + \left(4\tbeta_2+6\tbeta_2^2+4\tbeta_2^3+\tbeta_2^4\right)\tbeta_1 + 
(1+\talpha_1)\talpha_2^4 + \tgamma_1\tgamma_2^4 =0\cr
& \tbeta_2+ 2\tbeta_2^2 + 2\tbeta_2^3 + \tbeta_2^4 + {1\over 5}\left(
\talpha_2^5 + \tbeta_2^5 + \tgamma_2^5\right)=0.\cr}}
We denote by $H_1, \ldots, H_6$ the polynomials defined by the left hand side of the 
above equations so that $\CI=(H_1,\ldots, H_6)$ is 
the defining ideal of the Hilbert scheme in $U$.
The local ring of the Hilbert scheme at the point $F$ is the localization 
of the quotient ring 
$$ \CQ = \IC[\talpha_1, \tbeta_1,\tgamma_1,\talpha_2,\tbeta_2,\tgamma_2]/\CI$$ 
with respect to the maximal ideal $m_F\subset \CQ$. The later is generated by the images 
of $\talpha_1, \tbeta_1,\tgamma_1,\talpha_2,\tbeta_2,\tgamma_2$ in $\CQ$, therefore 
$\CQ_{m_F}$ is isomorphic to 
the ring of fractions of the form ${f\over g}$, where $g\in \CQ\setminus m_F$. 
The local analytic ring of the variety \hilbK\ at $F$ is given by the $m_F$-adic completion 
of $\CQ_{m_F}$. This is isomorphic to the quotient ring 
$$ 
\IC[[\talpha_1, \tbeta_1,\tgamma_1,\talpha_2,\tbeta_2,\tgamma_2]] / {\widehat \CI}
$$ 
where $\IC[[\talpha_1, \tbeta_1,\tgamma_1,\talpha_2,\tbeta_2,\tgamma_2]]$ is the ring of 
formal power series in six variables and ${\widehat \CI}$ is the ideal generated in it by the 
polynomials \hilbK.\
Let us perform the coordinate change
\eqn\hilbKi{\eqalign{
&\phi_1=\ta_1+2\ta_1^2+2\ta_1^3+\ta_1^4+{1\over 5}(\ta_1^5+\tb_1^5),\cr
&\phi_2=\ta_2+(4\ta_1+6\ta_1^2+4\ta_1^3+\ta_1^4)\ta_2+\tb_1^4(1+\tb_2),\cr
&\rho_1=\tb_1+(4\tb_2+6\tb_2^2+4\tb_2^3+\tb_2^4)\tb_1+\ta_2^4(1+\ta_1),\cr
&\rho_2=\tb_2+2\tb_2^2+2\tb_2^3+\tb_2^4+{1\over 5}(\ta_2^5+\tb_2^5),\cr
&\psi_1=\tg_1,~~\psi_2=\tg_2.\cr
}}
This is a nonsingular coordinate change in the ring of power series because its Jacobian matrix is 
nonsingular at the origin. Up to degree five, the inverse transformation reads
\eqn\hilbKii{\eqalign{
&\ta_1=\phi_1-2\phi_1^2+6\phi_1^3-21\phi_1^4+{399\over 5}\phi_1^5-{1\over 5}\rho_1^5+\ldots,\cr
&\ta_2=\phi_2-4\phi_1\phi_2+18\phi_1^2\phi_2-84\phi_1^3\phi_2-\rho_1^4+399\phi_1^4\phi_2
+4\phi_1\rho_1^4+15\rho_1^4\rho_2+\ldots,\cr
&\tb_1=\rho_1-4\rho_1\rho_2+18\rho_1\rho_2^2-\phi_2^4-84\rho_1\rho_2^3+399\rho_1\rho_2^4+4\phi_2^4
\rho_2+15\phi_1\phi_2^4+\ldots,\cr
&\tb_2=\rho_2-2\rho_2^2+6\rho_2^3-21\rho_2^4+{399\over 5}\rho_2^5-{1\over 5}\phi_2^5+\ldots~.\cr
}}
We can rewrite the first and last two equations in \hilbK\ as
\eqn\hilbKiii{\eqalign{
&\phi_1+{1\over 5}\psi_1^5=0,~~\phi_2+\psi_1^4\psi_2=0,\cr
&\rho_2+{1\over 5}\psi_2^5=0,~~\rho_1+\psi_1\psi_2^4=0.\cr
}}
Substituting these in the remaining two equations in \hilbK\ we obtain the equations describing the local analytic 
structure of the moduli space
\eqn\hilbKiV{\eqalign{
&\psi_1^3\psi_2^2+\psi_1^8\psi_2^2+\psi_1^{13}\psi_2^2-\psi_1^3\psi_2^{12}+\psi_1^{18}\psi_2^2
-2\psi_1^3\psi_2^{17}+\psi_1^{23}\psi_2^2+2\psi_1^8\psi_2^{17}-3\psi_1^3\psi_2^{22}+...=0,\cr
&\psi_1^2\psi_2^3+\psi_1^2\psi_2^8-\psi_1^{12}\psi_2^3+\psi_1^2\psi_2^{13}-2\psi_1^{17}\psi_2^3+\psi_1^2
\psi_2^{18}-3\psi_1^{22}\psi_2^3+2\psi_1^{17}\psi_2^8+\psi_1^2\psi_2^{23}+...=0.\cr
}}

{\it b) Deformed Case} 

The above computation can be fairly easily extended to the relative Hilbert scheme 
of lines associated to a family of quintic hypersurfaces. 
Let $\CX$ be such an $m$-parameter family of parameterized by 
$T=\IC^m$. The defining equations of $\CX$ in $\IP^4\times T$ are of the form 
\eqn\hilbA{ 
x_0^5+x_1^5+\ldots + x_4^5 + t_1G_1(x)+\ldots + t_m G_m(x) =0.}
Consider the curve $F\subset \CX$ determined by the equations 
\eqn\famhilbA{ 
x_0-\eta_{01}x_1=0,\qquad x_2-\eta_{23}x_3 =0, \qquad x_4=0, \qquad t_1=\ldots=t_m=0.}
Obviously, $F$ is embedded in the central fiber $\CX_0$. 
We would like to determine the analytic type at $F$ of the relative Hilbert scheme of lines 
for the family \hilbA.\ Since the parameter space is a linear space, this is equivalent to 
finding the local analytic structure of the Hilbert scheme of lines in the total space 
$\CX$ at $F$. 

We will proceed by analogy with the previous case. The Hilbert scheme in
question can now be represented as a complete intersection in $G(2,5) \times T$. 
Its local defining equations in $U\times T$ follow from a condition of the form  
\eqn\famhilbB{ \eqalign{
& -u^5+(\alpha_1 u + \alpha_2 v)^5 - v^5 +(\beta_1 u + \beta_2 v)^5 + (\gamma_1 u 
+ \gamma_2 v)^5+ K(t_1,\ldots, t_m, u,v)=0,\cr}}
for any values of $(u,v)\in \IC^2$. 
Here $K(t_1,\ldots, t_m, u,v)$ is a polynomial obtained by substituting 
\hilbG\ in the perturbation $G=t_1G_1(x)+\ldots +t_mG_m(x)$. 
By expanding all terms in \famhilbB,\ we obtain 
a system of polynomial equations which determine the ideal of the Hilbert scheme 
in $\IC[\talpha_1, \tbeta_1,\tgamma_1,\talpha_2,\tbeta_2,\tgamma_2]\times T$. 

For concreteness let us consider a one-parameter family defined by the perturbation $G=x_0x_2^2x_4^2$. 
Then the local equations of the relative Hilbert scheme in the neighborhood $U$ become
\eqn\hilpp{\eqalign{
&\ta_1+2\ta_1^2+2\ta_1^3+\ta_1^4+{1\over 5}(\ta_1^5+\tb_1^5+\tg_1^5)=0,\cr
&\ta_2+(4\ta_1+6\ta_1^2+4\ta_1^3+\ta_1^4)\ta_2+\tb_1^4(1+\tb_2)+\tg_1^4\tg_2=0,\cr
&\tb_1+(4\tb_2+6\tb_2^2+4\tb_2^3+\tb_2^4)\tb_1+\ta_2^4(1+\ta_1)+\tg_1\tg_2^4-{t\over 5}\tg_2^2=0,\cr
&\tb_2+2\tb_2^2+2\tb_2^3+\tb_2^4+{1\over 5}(\ta_2^5+\tb_2^5+\tg_2^5)=0,\cr
&(1+\ta_1)^3\ta_2^2+\tb_1^3(1+\tb_2)^2+\tg_1^3\tg_2^2-{t\over 10}\tg_1^2=0,\cr
&(1+\ta_1)^2\ta_2^3+\tb_1^2(1+\tb_2)^3+\tg_1^2\tg_2^3-{t\over 5}\tg_1\tg_2=0.\cr
}}
We perform the same coordinate change as described in \hilbKi\ and rewrite the first four equations in 
\hilpp\ as 
\eqn\hilppi{\eqalign{
&\phi_1+{1\over 5}\psi_1^5=0,~~\phi_2+\psi_1^4\psi_2=0,\cr
&\rho_2+{1\over 5}\psi_2^5=0,~~\rho_1-{t\over 5}\psi_2^2+\psi_1\psi_2^4=0.\cr
}}
Substituting now in the last two equations in \hilpp\ and using \hilbKii\ we obtain the following equations
\eqn\hilppii{\eqalign{
&-{t\over 10}\psi_1^2+\psi_1^3\psi_2^2+{t^3\over 125}\psi_2^6-{3t^2\over 25}\psi_1
\psi_2^8+\psi_1^8\psi_2^2+\ldots =0,\cr
&-{t\over 5}\psi_1\psi_2+{t^2\over 25}\psi_2^4+\psi_1^2\psi_2^3-{2t\over 5}\psi_1
\psi_2^6+{t^2\over 25}\psi_2^9+\psi_1^2\psi_2^8+\ldots =0.\cr
}}

\subsec{Homotopy Transformations -- Landau-Ginzburg Phase} 

Let us recall our results for higher products in Landau-Ginzburg phase. For simplicity, we shall denote $H^k=H^k(\of,\of),\ k=0,1$ as in section $3$.

{\it a) Undeformed Case} 

After a trivial rescaling, we have 
\eqn\prodA{\eqalign{ 
& m_5(\Phi^{\otimes 5}) = \psi_1^3\psi_2^2 U + \psi_1^2\psi_2^3 V\cr
& m_{15}(\Phi^{\otimes 15}) = \left({3\over 100}\psi_1^{13}\psi_2^2+{1\over 4}\psi_1^3\psi_2^{12}\right)U
+\left({3\over 100}\psi_2^{13}\psi_1^2+{1\over 4}\psi_2^3\psi_1^{12}\right)V.\cr}}
We would like to prove that all products $m_n$, $n>5$ can be set to zero by performing 
homotopy transformations. That is the $A_\infty$ structure \prodA\ is homotopic equivalent to 
a new structure defined by 
$m'_n:H^{\otimes n} \ra H$, $n\geq 1$ so that 
\eqn\prodB{ 
m'_n(\Phi^{\otimes n}) = \left\{\matrix{m_5,\hfill& \hbox{if}\ n=5\cr
0,\hfill &\hbox{otherwise}\cr}\right.}
Since we do not have closed formulae for all the products $m_n$, we will only check this 
claim up to degree fifteen. Recall that a homotopy transformation between two $A_\infty$ structures 
is given by a sequence of linear maps $f_n:H^{\otimes n}\ra H$ of degree $1-n$, $f_1=\II$, so that 
\eqn\prodC{ 
\sum (-1)^{r+st}f_u\left(\II^{\otimes r} \otimes m_s'\otimes \II^{\otimes t}\right) 
= \sum (-1)^\sigma m_r \left(f_{i_1}\otimes f_{i_2} \otimes f_{i_2} \otimes 
\cdots \otimes
f_{i_r}\right).}
The sum in the left hand side of \prodC\ runs over all decompositions $n=r+s+t$, 
and $u=r+t+1$. The sum in the right hand side runs over all $1\leq r \leq n$ and 
all decompositions 
$n=i_1+\cdots + i_r$, and $\sigma = (r-1)(i_1-1) +(r-2)(i_2-1) + \cdots + 
2(i_{r-2}-1)+(i_{r-1}-1)$.
In addition, when applying these formulae to elements, we have to take into account 
the Koszul sign rule 
\eqn\prodD{(f\otimes g)(x\otimes y) = (-1)^{{\tilde g}{\tilde x}} f(x) \otimes g(y),}
where ${\tilde g}$, ${\tilde x}$ denote the degrees of $g$ and respectively $x$.

For a more transparent understanding of the homotopy transformations \prodC\ let us study their 
action on $n$-uples of the form $\left(\psi_1 S + \psi_2 T,\ldots, \psi_1 S + \psi_2 T\right)$. 
Note that $f_{i_s}\left(\psi_1 S + \psi_2 T,\ldots, \psi_1 S + \psi_2 T\right)$ takes values in odd cohomology since all arguments are odd and the degree of $f_{i_s}$ is $1-i_s$. Taking into account 
the signs, it follows that the right hand side of \prodC\ represents the effect of a 
field redefinition of the form 
\eqn\prodD{ 
\Phi\,\, \ra\,\, \Phi+f_2(\Phi,\Phi) +\ldots + f_n(\Phi,\ldots, \Phi) + \ldots }
on the products $m_n(\Phi,\ldots, \Phi)$. 
To interpret the terms in the left hand side, recall that the moduli space is defined by 
\eqn\prodF{ 
\sum_{n\geq 1} (-1)^{n(n+1)/2}m_n(\Phi^{\otimes n}) =0.}
Moreover, homotopic equivalent $A_\infty$ structures should produce isomorphic moduli spaces.
As explained above, the terms in the right hand side take into account the effect of automorphisms 
of the ring of formal power series $\IC[[\psi_1,\psi_2]]$. Apart from ring automorphisms, the 
analytic type of the moduli space should also be invariant under a change of generators 
in the ideal defined by \prodF.\ Such transformations on ideal generators
are encoded in the left hand side of \prodC.\ 

In the following we will show that in the undeformed case the higher products can be set 
in the form \prodB\ up to degree fifteen using only field redefinitions of the form \prodD.\ 
Let us write 
\eqn\prodG{ 
f_n\left((\psi_1 S+ \psi_2 T)^{\otimes n}\right) =
p_n(\psi_1,\psi_2)S+q_n(\psi_1,\psi_2)T,\ n\geq 2}
where $p_n(\psi_1,\psi_2), q_n(\psi_1,\psi_2)$ are arbitrary homogeneous polynomials in $(\psi_1,\psi_2)$ of degree $n$. We also assume that the $f_n$ act trivially on even elements so that 
\prodC\ yields
\eqn\prodH{ 
m'_n\left((\psi_1 S+ \psi_2 T)^{\otimes n}\right)= 
\sum m_r\left(f_{i_1}((\psi_1 S+ \psi_2 T)^{\otimes i_1}),\ldots 
f_{i_r}((\psi_1 S+ \psi_2 T)^{\otimes i_1})\right).}
Taking into account the sign rule \prodD,\ all terms in the right hand side of equation \prodC\ have sign $+$ when evaluated on fermionic elements. 

Then some straightforward linear algebra shows that we can choose $p_n=q_n=0$ for $2\leq n\leq 10$. 
Setting $m'_{15}$ to zero, we obtain the following linear equations for $p_{11},q_{11}$ 
\eqn\prodI{\eqalign{
& 3 \psi_1^2\psi_2^2 p_{11} + 2 \psi_1^3\psi_2 q_{11} = -\left({3\over 100}\psi_1^{13}\psi_2^2+{1\over 4}\psi_1^3\psi_2^{12}\right)\cr
& 2 \psi_1\psi_2^3 p_{11} +  3 \psi_1^2\psi_2^2 q_{11} = -\left({3\over 100}\psi_2^{13}\psi_1^2+
{1\over 4}\psi_2^3\psi_1^{12}\right)\cr}}
which has the unique solution 
\eqn\prodJ{ 
p_{11}(\psi_1,\psi_2) = {\psi_1\over 500} (41\psi_1^{10}-69\psi_2^{10}),\qquad 
q_{11}(\psi_1,\psi_2) = {\psi_2\over 500} (41\psi_2^{10}-69\psi_1^{10}).}
Proceeding similarly, we can in principle set all the higher order products to zero unless 
at some order, the resulting linear system has no polynomial solutions. We do not know if such an obstruction arises, but we conjecture that it is absent. 

{\it b) Deformed Case} 

We rescale again the $A_\infty$ coefficients by a trivial constant factor, and we also 
set $\eta=-1$ for simplicity. Then we obtain
\eqn\defA{\eqalign{
m_2^\alpha(\Phi^{\otimes 2})&= 
-{\alpha\over 10}\Big(\psi_1^2 U+2\psi_1\,\psi_2 V\Big)\cr
m_4^\alpha(\Phi^{\otimes 4}) &=
{\alpha^2\over 50}\Big(2\psi_2^3\,\psi_1 U+\psi_2^4 V\Big)\cr 
m_5^\alpha(\Phi^{\otimes 5}) &= \psi_1^3\psi_2^2 U + \psi_1^2\psi_2^3 V \cr
m_6^\alpha(\Phi^{\otimes 6}) &= {\alpha^3\over 1000} 
\Big(\psi_2^6U+2\psi_2^2\psi_1^4 V\Big).\cr}}
In this case it is straightforward to check that any homotopy transformation will leave 
the first two products unchanged, that is 
\eqn\defB{ 
{m'}_1^\alpha (\Phi) = m_1^\alpha (\Phi) = 0, \qquad 
{m'}_2^\alpha (\Phi^{\otimes 2}) = {m}_2^\alpha (\Phi^{\otimes 2}) =-{\alpha\over 10}\Big(\psi_1^2 U + 2\psi_1\,\psi_2 V\Big).}Next, we will try to set all higher order products except $m_5^\alpha(\Phi^{\otimes \alpha})$  to zero by analogy with the undeformed case.  
The third product is zero, so we do not have to perform any transformation in degree two i.e.  $f_2=0$. The fourth product is more interesting. Making equations \prodC\ explicit and using \defB,\ we obtain 
\eqn\defC{ \eqalign{ 
& {m'}_4^\alpha(\Phi,\Phi,\Phi,\Phi) + f_3(m_2^\alpha(\Phi,\Phi),\Phi,\Phi)-
f_3(\Phi,m_2^\alpha(\Phi,\Phi),\Phi) + f_3(\Phi,\Phi,m_2^\alpha(\Phi,\Phi))=\cr
& m_4^\alpha(\Phi,\Phi,\Phi,\Phi) + m_2^\alpha(f_3(\Phi,\Phi,\Phi),\Phi) + 
m_2^\alpha(\Phi,f_3(\Phi,\Phi,\Phi)).\cr}}
Writing again 
\eqn\defD{ 
f_3\left((\psi_1 S+ \psi_2 T)^{\otimes n}\right)=p_3(\psi_1,\psi_2)S+q_3(\psi_1,\psi_2)T}
we can evaluate 
\eqn\defEA{
m_2^\alpha(f_3(\Phi,\Phi,\Phi),\Phi) + m_2^\alpha(\Phi,f_3(\Phi,\Phi,\Phi))
= -{\alpha\over 5} p_3\psi_1 U -{\alpha\over 5} p_3\psi_2 V - {\alpha\over 5} q_3 \psi_1 V.}
Then the right hand side of equation \defC\ becomes 
\eqn\defE{ 
\left({\alpha^2\over 25}\psi_2^3\psi_1 -{\alpha\over 5} p_3\psi_1\right) U + 
\left({\alpha^2\over 50} \psi_2^4 -{\alpha \over 5}p_3\psi_2 -{\alpha\over 5} q_3\psi_1 \right)V\,,}
while the left hand side reads  
\eqn\defF{\eqalign{ 
{m'}_4^\alpha(\Phi,\Phi,\Phi,\Phi)+ f_3\left(-{\alpha\over 10} \psi_1^2 U -{\alpha \over 5}\psi_1\psi_2 V\,, \psi_1 S +\psi_2 T\,, \psi_1 S + \psi_2 T\right) 
+ \hbox{permutations}}\,.}
Note that the term  $-{\alpha\over 5} q_3\psi_1 V$ that appears in \defEA\ can be absorbed in the definition of $f_3(m_\alpha^2(\Phi,\Phi),\Phi,\Phi)$ in \defF\ and we can set $q_3=0$. 
Now it is clear that in order to eliminate the term ${\alpha^2\over 50}\,\psi_2^4 V$ in 
$m_4^\alpha(\Phi^{\otimes 4})$ we have to choose
\eqn\defG{ 
p_3 ={\alpha\over 10} \psi_2^3.} 
Then formula \defE\ reduces to 
\eqn\defH{ 
{\alpha^2\over 50} \psi_2^3\psi_1 U.} 
This term can be removed by a judicious choice of $f_3$ in \defF\ and we can set 
${m'}_4^\alpha(\Phi^{\otimes 4})=0$. 

Proceeding similarlgy, it is straightforward to maintain 
${m'}_5^\alpha({\Phi}^{\otimes 5})=m_5^\alpha(\Phi^{\otimes 5})$
by choosing $f_4=0$. However, the degree six product is more interesting. Taking into account $f_2=f_4=0$, the right hand side of \prodC\ bcomes 
\eqn\defI{ \eqalign{
& m^\alpha_6(\Phi^{\otimes 6}) + m_4^\alpha(\Phi^{\otimes 3}, f_3(\Phi^{\otimes 3})) + m_4^\alpha(\Phi,f_3(\Phi^{\otimes 3}),\Phi^{\otimes 2}) + \ldots \cr
& + m_4^\alpha(f_3(\Phi^{\otimes 3}),\Phi^{\otimes 3})+ m_2^\alpha(f_3(\Phi^{\otimes 3}),f_3(\Phi^{\otimes 3})) = 
\left({\alpha^3\over 250}\psi_2^6 U + {\alpha^3\over 500}\psi_1^4\psi_2^2 V\right).\cr}}
The left hand side of \prodC\ reduces to 
\eqn\defJ{\eqalign{
& {m'}^\alpha_6(\Phi^{\otimes 6})+ \sum_{r+t=4} (-1)^rf_5\left(\Phi^{\otimes r}\otimes m_2^\alpha(\Phi,\Phi), \Phi^{\otimes t}\right)=\cr
& {m'}^\alpha_6(\Phi^{\otimes 6})+
\sum_{r+t=4} f_5\left(\Phi^{\otimes r}, -{\alpha\over 10}\psi_1^2 U -{\alpha\over 5}\psi_1\psi_2 V, 
\Phi^{\otimes t}\right).\cr}}
It is clear that we can remove the term ${\alpha^3\over 500} \psi_1^4\psi_2^2 V$
from \defI\ by a judicious choice of $f_5$. However, since $f_5$ is linear we cannot change the term ${\alpha^3\over 250}\psi_2^6 U$ because the coefficient of $\psi_2^6$ in \defJ\ is zero. This is an obstruction preventing us from setting ${m'}^\alpha_6(\Phi^{\otimes 6})$ to $0$. 
Therefore, up to order six, the deformed $A_\infty$ structure \defA\ can be set in the form 
\eqn\defK{\eqalign{ 
& {m'}_2^\alpha(\Phi^{\otimes 2}) = -{\alpha\over 10}\Big(\psi_1^2U+2\psi_1\,\psi_2 V\Big)\cr
& {m'}_5^\alpha(\Phi^{\otimes 5}) = \left(\psi_1^3\psi_2^2 U + \psi_1^2\psi_2^3 V\right))\cr 
& {m'}_6^\alpha(\Phi^{\otimes 6}) = {\alpha^3\over 250}\psi_2^6 U.\cr}}
In principle we can carry out this algorithm up to arbitrarily high order. 
Note that a priori there is no canonical form for the $A_\infty$ coefficients. For example, by making different homotopy transformations one can set the $A_\infty$ structure in the form 
\eqn\defL{\eqalign{ 
& {m'}_2^\alpha (\Phi^{\otimes 2}) = -{\alpha\over 10}\Big(\psi_1^2U+2\psi_1\,\psi_2 V\Big)\cr
& {m'}_4^\alpha(\Phi^{\otimes 4}) = {\alpha^2\over 50}\left(4\psi_1\psi_2^3 U + \psi_2^4 V\right)\cr
& {m'}_5^\alpha(\Phi^{\otimes 5}) = \left(\psi_1^3\psi_2^2 U + \psi_1^2\psi_2^3 V\right))\cr 
& {m'}_6^\alpha(\Phi^{\otimes 6}) = 0.\cr}}
However the important lesson we should draw from this computation is that the higher products contain non homotopically trivial 
$\alpha$-dependent terms beyond the leading order $n=2$. Presumably, this information should be 
properly encoded in certain homotopy invariants of the $A_\infty$ structure. We will leave this 
question for future work. 

Note also that in the original form the moduli space \prodF\ cannot be represented as the critical locus 
of a D-brane superpotential $W(\psi_1,\psi_2)$. However the moduli space associated to the $A_\infty$ 
structure \defK\ can be written as the critical locus of 
\eqn\defM{ 
W(\psi_1,\psi_2) ={1\over 3} \psi_1^3\psi_2^3 -{\alpha\over 10} \psi_1^2\psi_2 + {\alpha^3\over 1750} \psi_2^7+\ldots}
at least up to order seven. Alternatively, using the products \defL\ instead we find \eqn\defN{ 
W(\psi_1,\psi_2) ={1\over 3}\psi_1^3\psi_2^3 -{\alpha\over 10} \psi_1^2\psi_2-{\alpha^2\over 50}\psi_1\psi_2^4+\ldots.}
Note that the leading term in this expression was conjectured in \BDLR\ starting from enumerative considerations. See also \refs{\CK,\versal}. Here we have derived it 
from Landau-Ginzburg considerations together with the first two homotopically 
nontrivial closed string corrections.

The fact that the superpotential is only uniquely defined up to homotopy transformations may seem puzzling from a physical point of view. However, recall that so far we have been working exclusively in the framework of topological string theory. In a full fledged superstring theory, we would also have to specify a kinetic term and a measure on the space of massless fields. Usually most physical situations would require a specific canonical form for these
quantities, in which case the superpotential will also be uniquely determined\foot{We thank Mike Douglas for a very useful discussion on this point.}. We hope to address this problem elsewhere. 

To conclude this subsection, note that the above homotopy transformations are equivalent to a change of basis in the moduli space ideal \prodF\ ffollowed by an invertible coordinate transformation. Let us write the equations \prodF\ in the form $e_1=e_2=0$ where 
\eqn\defO{\eqalign{ 
& e_1 = \psi_1^3\psi_2^2 -{\alpha\over 10} \psi_1^2 +{\alpha^2\over 25} \psi_2^3\psi_1 +{\alpha^3\over 1000}\psi_2^6+\ldots  \cr
& e_2 = \psi_2^3\psi_1^2 -{\alpha \over 5} \psi_1\psi_2 +{\alpha^2\over 50} \psi_2^4 + {\alpha^3\over 500} \psi_2^2\psi_1^4+\ldots~. \cr}}
Now perform the following transformation 
\eqn\defP{\eqalign{ 
& e'_1=e_1+{\alpha \over 10} \psi_2^2 e_2 \cr
& e'_2=e_2+{\alpha^2\over 50} \psi_1^2\psi_2^2 e_1.\cr}} 
We obtain 
\eqn\defQ{\eqalign{ 
& e_1'= \psi_1^3\psi_2^2 -{\alpha\over 10} \psi_1^2 +{\alpha^2\over 50} \psi_2^3\psi_1 +{3 \alpha^3\over 1000} \psi_2^6+\ldots  \cr
& e_2'= \psi_2^3\psi_1^2 -{\alpha \over 5} \psi_1\psi_2 +{\alpha^2\over 50} \psi_2^4 +\ldots~. \cr}}
Next, after change of variables of the form 
\eqn\defR{
\psi_1 = \tpsi_1 +{\alpha \over 10} \tpsi_2^3 + \ldots \qquad 
\psi_2 = \tpsi_2 +\ldots}
formula \defQ\ becomes 
\eqn\defR{\eqalign{ 
& e_1' = \tpsi_1^3\tpsi_2^2 -{\alpha \over 10} \tpsi_1^2 +{\alpha^3 \over 250} \tpsi_2^6+\ldots \cr
& e'_2 =\tpsi_2^3\tpsi_1^2 -{\alpha \over 5} \tpsi_1\tpsi_2 +\ldots~. \cr}}
This is in agreement with \defK.\ 
In the following we will carry out a similar algorithm for the moduli 
space equations in the geometric phase, and show that they agree with the Landau-Ginzburg results up to homotopy transformations.

\subsec{Homotopy Transformations -- Calabi-Yau Phase} 

{\it a) Undeformed Case}

Proceeding as above, let us write the equations \hilbKiV\ giving the generators of the moduli 
space ideal in the form $e_1=e_2=0$, where
$$
e_1=\psi_1^3\psi_2^2+\psi_1^8\psi_2^2+\psi_1^{13}\psi_2^2-\psi_1^3\psi_2^{12}+\psi_1^{18}\psi_2^2
-2\psi_1^3\psi_2^{17}+\psi_1^{23}\psi_2^2+2\psi_1^8\psi_2^{17}-3\psi_1^3\psi_2^{22}+\ldots
$$
$$
e_2=\psi_1^2\psi_2^3+\psi_1^2\psi_2^8-\psi_1^{12}\psi_2^3+\psi_1^2\psi_2^{13}-2\psi_1^{17}\psi_2^3+\psi_1^2
\psi_2^{18}-3\psi_1^{22}\psi_2^3+2\psi_1^{17}\psi_2^8+\psi_1^2\psi_2^{23}+\ldots~.
$$
Let us perform now the following coordinate transformation
\eqn\coordK{\eqalign{
\psi_1=&\tpsi_1-{3\over 5}\tpsi_1^6+{2\over 5}\tpsi_1\tpsi_2^5+{32\over 25}\tpsi_1^{11}-{16\over 25}
\tpsi_1^6\tpsi_2^5-{8\over 25}\tpsi_1\tpsi_2^{10}-{437\over 125}\tpsi_1^{16}+{394\over 125}\tpsi_1^{11}
\tpsi_2^5-{71\over 125}\tpsi_1^6\tpsi_2^{10}\cr
&+{48\over 125}\tpsi_1\tpsi_2^{15}+{6649\over 625}\tpsi_1^{21}-{7844\over 625}\tpsi_1^{16}\tpsi_2^5+
{1679\over 625}\tpsi_1^{11}\tpsi_2^{10}+{286\over 625}\tpsi_1^6\tpsi_2^{15}-{211\over 625}\tpsi_1
\tpsi_2^{20},\cr
\psi_2=&\tpsi_2+{2\over 5}\tpsi_1^5\tpsi_2-{3\over 5}\tpsi_2^6-{8\over 25}\tpsi_1^{10}\tpsi_2
-{16\over 25}\tpsi_1^5\tpsi_2^6+{32\over 25}\tpsi_2^{11}+{48\over 125}\tpsi_1^{15}\tpsi_2
-{71\over 125}\tpsi_1^{10}\tpsi_2^6+{394\over 125}\tpsi_1^{5}
\tpsi_2^{11}\cr
&-{437\over 125}\tpsi_2^{16}-{211\over 625}\tpsi_1^{20}\tpsi_2+{286\over 625}\tpsi_1^{15}\tpsi_2^6
+{1679\over 625}\tpsi_1^{10}\tpsi_2^{11}-{7844\over 625}\tpsi_1^{5}\tpsi_2^{16}
+{6649\over 625}\tpsi_2^{21}.
}}
We now obtain
\eqn\coordKi{\eqalign{
&e_1=\tpsi_1^3\tpsi_2^2+\ldots\cr
&e_2=\tpsi_1^2\tpsi_2^3+\ldots~.}}
Although we can not prove that all the terms of degree higher vanish, we conjecture that that is true.

{\it b) Deformed Case}

We write now the equations \hilppii\ for the generators of the moduli space ideal in the deformed case 
as $e_1^d=e_2^d=0$, where
$$
e_1^d=-{t\over 10}\psi_1^2+\psi_1^3\psi_2^2+{t^3\over 125}\psi_2^6-{3t^2\over 25}\psi_1
\psi_2^8+\psi_1^8\psi_2^2+\ldots,
$$
$$
e_2^d=-{t\over 5}\psi_1\psi_2+{t^2\over 25}\psi_2^4+\psi_1^2\psi_2^3-{2t\over 5}\psi_1
\psi_2^6+{t^2\over 25}\psi_2^9+\psi_1^2\psi_2^8+\ldots~.
$$
We first perform a change of variables of the form
\eqn\dchangei{
\psi_1=\tpsi_1+{t\over 5}\tpsi_2^3,\qquad \psi_2=\tpsi_2.
}
We obtain
\eqn\changdegeni{\eqalign{
&e_1^d=-{t\over 10}\tpsi_1^2-{t^2\over 25}\tpsi_1\tpsi_2^3+\tpsi_1^3\tpsi_2^2+
{t^3\over 250}\tpsi_2^6+{3t\over 5}\tpsi_1^2\tpsi_2^5+\tpsi_1^8\tpsi_2^2+\ldots,\cr
&e_2^d=-{t\over 5}\tpsi_1\tpsi_2+\tpsi_1^2\tpsi_2^3+\tpsi_1^2\tpsi_2^8+\ldots~.
}}
Next, we perform the following change of generators
\eqn\changegenii{
e_1^{d'}=e_1^d-{t\over 5}\tpsi_2^2e_2^d,\qquad e_2^{d'}=e_2^d.
}
The new generators are given by
\eqn\changedgeniii{\eqalign{
&e_1^{d'}=-{t\over 10}\tpsi_1^2+\tpsi_1^3\tpsi_2^2+
{t^3\over 250}\tpsi_2^6+{3t\over 5}\tpsi_1^2\tpsi_2^5+\tpsi_1^8\tpsi_2^2+\ldots,\cr
&e_2^{d'}=-{t\over 5}\tpsi_1\tpsi_2+\tpsi_1^2\tpsi_2^3+\tpsi_1^2\tpsi_2^8+\ldots~.}}
We now perform the coordinate change
\eqn\cchangedefii{\eqalign{
&\tpsi_1={\tPsi_1}+2{\tPsi_1}{\tPsi_2}^5+{5\over t}{\tPsi_1}^7{\tPsi_2}^2+{10\over t}{\tPsi_1}^2
{\tPsi_2}^7,\cr
&\tpsi_2=\tPsi_2-2\tPsi_2^6-{5\over t}\tPsi_1^6\tPsi_2^3-{15\over t}\tPsi_1\tPsi_2^8
}}
and obtain the following expressions for the generators
\eqn\changedgenV{\eqalign{
&e_1^{d'}=-{t\over 10}\tPsi_1^2+\tPsi_1^3\tPsi_2^2+{t^3\over 250}\tPsi_2^6+\ldots,\cr
&e_2^{d'}=-{t\over 5}\tPsi_1\tPsi_2+\tPsi_1^2\tPsi_2^3+\ldots~.
}}
This result is in agreement with \defR,\ therefore providing strong evidence for the equivalence 
of the two approaches.

\appendix{A}{Horizontal Deformations in Landau-Ginzburg Families}

In this appendix we present a detailed derivation of 
the odd morphism space $H^1(\OCF,\OCF)$, which plays an important role in section four. 
For simplicity we consider a one parameter deformation 
$W_{LG}+tG$ of the Landau-Ginzburg model. 
Recall that $\OCF$ denotes a tensor product factorization of the deformed superpotential 
of the form 
\eqn\appA{
\OCF= \of\otimes 
\left(\xymatrix{\IC[x_i,t] \ar@<1ex>[r]^{t}& \IC[x_i,t]\ar@<1ex>[l]^{G}\\}\right)}
where $t$ is regarded as a dynamical variable. Let us denote the second factor in 
the right hand side of \appA\ by $\om_t$. We will also write the factorization $\of$ 
of $W_{LG}$ in the form 
\eqn\appB{ 
\of = \left(\xymatrix{F_1 \ar@<1ex>[r]^{f_1}& F_0\ar@<1ex>[l]^{f_0}\\}\right)}
where $F_0, F_1$ are free $R$-modules of equal rank and $f_0, f_1$ are $R$-module 
homomorphisms so that $f_1f_0=f_1f_0=W_{LG}$. Recall that $R=\IC[x_0,\ldots, x_n]$. 
The tensor product factorization takes the form 
\eqn\appC{ 
\OCF = \left(\xymatrix{\CF_1 \ar@<1ex>[r]^{\phi_1}& \CF_0\ar@<1ex>[l]^{\phi_0}\\}\right)}
where $\CF_1, \CF_0$, $\phi_1, \phi_0$ are $R[t]$-modules and respectively $R[t]$-module 
homomorphisms. Using the general formulae \tensorA\ and \tensorB\ we have 
\eqn\appD{\eqalign{ 
& \CF_1 = F_1\otimes_R R[t] \oplus F_0\otimes_R R[t] \cr
& \CF_0 = F_0\otimes_R R[t] \oplus F_1\otimes_R R[t].\cr}}
The maps $\phi_1, \phi_0$ can be written as block matrices 
\eqn\appE{ 
\phi_1 = \left[\matrix{G & f_0\cr f_1 & -t\cr}\right],\qquad 
\phi_0=\left[\matrix{t & f_0 \cr f_1 & -G\cr}\right]}
with respect to the direct sum decomposition \appD.\ 

The cochains in the morphism complex $\IH(\OCF,\OCF)$ can be similarly written in block form. 
Even cochains are pairs of $R[t]$-module homomorphisms $A:\CF_1\ra \CF_1$, 
$B:\CF_0\ra \CF_0$ which can be written as  
\eqn\appF{ 
A = \left[\matrix{ A_{11} & A_{10} \cr A_{01} & A_{00}\cr}\right],\qquad 
B = \left[\matrix{ B_{11} & B_{10} \cr B_{01} & B_{00}\cr}\right].}
Odd cochains are pairs $(T,S)$, $T:\CF_1\ra \CF_0$, $S:\CF_0\ra \CF_1$ of the form 
\eqn\appG{ 
T:= \left[\matrix{ T_{11} & T_{10} \cr T_{01} & T_{00}\cr}\right],\qquad 
S:= \left[\matrix{ S_{11} & S_{10} \cr S_{01} & S_{00}\cr}\right].}
The differential $\CD$ in the deformed theory is given by 
\eqn\appG{\eqalign{
& \CD(A,B)=(A\phi_0-\phi_0 B, -\phi_1 A + B \phi_1)\cr
& \CD(T,S)=(\phi_0 T+S\phi_1, T\phi_0+\phi_1 S).\cr}}
We want to determine the odd cohomology of this complex. We can use the equivalence relation 
$(T,S) \sim (T',S') \Leftrightarrow (T',S')-(T,S) \in \hbox{Im}(\CD)$ to set $(T,S)$ in a
special form. Namely we can take $S_{11}, S_{01}, T_{10}, T_{00}$ to be independent of $t$. 
Note that this is not a single valued parameterization of the 
coset space $\IH^1(\OCF,\OCF)/\hbox{Im}(\CD)$. The cochains of this form are still subject to 
residual equivalence relations which will be made more explicit below. 

Substituting 
\appF\ and \appG\ in the $\CD$-closure condition $\CD(T,S)=0$ we find that 
\eqn\appH{ 
S_{00}=T_{11}=0} 
and 
\eqn\appI{
S_{01}=T_{01}, \qquad S_{10}=T_{10}} 
must also be independent on $T$. Moreover we have 
\eqn\appK{\eqalign{ 
& f_0T_{00}+S_{11}f_0=0\cr
& T_{00}f_1 + f_1 S_{11} =0\cr
& GS_{11} + f_0T_{01} + T_{10}f_1 =0 \cr
& -GT_{00} + f_1T_{10}+T_{01}f_0 =0.\cr}}
Regarding $\beta=(-S_{11},T_{00})$ and  $\xi=(T_{10}, T_{01})$ as cochains in the morphism complex 
$\IH(\of,\of)$, we can rewrite \appK\ as 
\eqn\appL{\eqalign{ 
& G(x_i)\,\beta=D\cdot\xi,\qquad D\cdot\beta =0.\cr}}
where $D$ is the differential of the undeformed theory. This is equation \cpxmodM\ in section four. The residual equivalence relations 
take the form 
\eqn\appM{\eqalign{ 
& \beta\sim \beta' \Leftrightarrow \beta'-\beta \in \hbox{Im}(D) \cr 
& \xi\sim \xi' \Leftrightarrow \xi'-\xi \in \hbox{Ker}(D).\cr}}
These are precisely the equivalence relations stated below \cpxmodM.\ 

Now suppose $\beta=\II$ and $G$ is trivializable, that is $G=D\cdot\xi$ for a fermionic cochain $\xi$. Then the pair $(\xi,\II)$ determines an odd cohomology representative $\Xi \in \IH^1(\OCF,\OCF)$ 
which reads 
\eqn\appN{ 
\Xi = \left(\left[\matrix{0 & T_{10}\cr T_{01} & -\II\cr}\right], 
\left[\matrix{\II & T_{10} \cr T_{01} & 0 \cr}\right]\right)}
in block form. As explained in section four, the equations of the local moduli space of 
$\OCF$ are determined by higher products evaluated on fermionic morphisms of the form 
$\Xi+\Lambda$, where $\Lambda \in \IH^1(\OCF,\OCF)$ is a cohomology representative 
in the image of the embedding 
\eqn\appO{
0\ra \IH^1(\of,\of){\buildrel \otimes \II\over \ra} \IH^1(\OCF,\OCF).}
Morphisms in the image of \appO\ can be written in block form as 
\eqn\appQ{ 
\Lambda = \left(\left[\matrix{0 & \mu \cr  \rho & 0 \cr}\right], \left[\matrix{ 
0 & \rho\cr \mu & 0 \cr}\right]\right)}
where $\Phi =(\mu, \rho)\in \IH^1(\of,\of)$. 
Then it is straightforward to check by matrix multiplication that all products of the form 
$\Xi\Lambda, \Lambda\Xi, \Xi^2, \Lambda^2$ lie in the image of the embedding map 
\eqn\appR{ 
0\ra \IH^0(\of,\of) {\buildrel \otimes \II\over \ra}\IH^0(\OCF,\OCF).}

\listrefs
\end